\documentclass[11pt,a4paper]{article}
\pdfoutput=1
\usepackage[applemac]{inputenc}
\usepackage{amsmath, amsthm}
\usepackage{jheppub}
\usepackage[center]{caption}
\usepackage{subcaption}
\usepackage{multirow}
\usepackage{booktabs}
\usepackage{amsmath}
\usepackage{amsfonts}
\usepackage{amssymb}
\usepackage{graphicx}
\usepackage{indentfirst}
\usepackage{slashed}
\usepackage{diagbox}
\usepackage{mathrsfs}
\usepackage{hyperref}
\usepackage{bm}
\usepackage{feynmp}
\usepackage{bbold}

\allowdisplaybreaks

\makeatletter
\newcommand\makebig[2]{%
  \@xp\newcommand\@xp*\csname#1\endcsname{\bBigg@{#2}}%
  \@xp\newcommand\@xp*\csname#1l\endcsname{\@xp\mathopen\csname#1\endcsname}%
  \@xp\newcommand\@xp*\csname#1r\endcsname{\@xp\mathclose\csname#1\endcsname}%
}
\makeatother

\makebig{biggg} {3.0}
\makebig{Biggg} {3.5}
\makebig{bigggg}{4.0}
\makebig{Bigggg}{4.5}

\newcommand{\cproj}[1]{\mathbb{P}^{#1}(\mathbb{C})}
\newcommand{\rproj}[1]{\mathbb{P}^{#1}(\mathbb{R})}

\newcommand{\refE}[1]{eq.~(\ref{#1})}

\renewcommand\Re{\operatorname{\mathfrak{Re}}}

\newcommand{\DeltaMPL}{\Delta}

\def\beq{\begin{equation}}
\def\eeq{\end{equation}}
\def\bsp#1\esp{\begin{split}#1\end{split}}
\newcommand{\bea}{\begin{eqnarray}}
\newcommand{\eea}{\end{eqnarray}}
\newcommand{\bean}{\begin{eqnarray*}}
\newcommand{\eean}{\end{eqnarray*}}

\def\det{\mathop{\rm det}}

\def\a{{\alpha}}
\def\b{{\beta}}
\def\g{{\gamma}}
\def\c{{\gamma}}

\def\eps{\epsilon}
\def\ep{\epsilon}

\def\ord{{\cal O}}

\def\Label#1{\label{#1}
  \smash{\hbox to0pt{\raise1ex\hbox{\tiny[#1]}\hss}}}

\def\beq{\begin{equation}}
\def\eeq{\end{equation}}
\def\bsp#1\esp{\begin{split}#1\end{split}}

\newcommand{\cC}{\mathcal{C}}

\def\vev#1{\left\langle #1 \right\rangle}

\renewcommand{\ln}{\log}

\theoremstyle{definition}

\DeclareMathOperator*{\res}{Res}

\preprint{CERN-TH-2019-168
\rightline{CP3-19-48}
}

\title{From positive geometries to a coaction on hypergeometric functions}

\author[a]{Samuel Abreu,}
\author[b,c,d]{Ruth Britto,}
\author[e]{Claude Duhr,}
\author[f]{Einan Gardi,}
\author[f]{and James Matthew}

\affiliation[a]{Center for Cosmology, Particle Physics and Phenomenology (CP3), Universit\'e Catholique de Louvain, 1348 Louvain-La-Neuve, Belgium}
\affiliation[b]{School of Mathematics, Trinity College, Dublin 2, Ireland}
\affiliation[c]{Hamilton Mathematics Institute, Trinity College, Dublin 2, Ireland}
\affiliation[d]{Institut de Physique Th{\'e}orique, Universit\'e Paris Saclay, 
CEA, CNRS, F-91191 Gif-sur-Yvette cedex, France}
\affiliation[e]{Theoretical Physics Department, CERN, Geneva, Switzerland}
\affiliation[f]{Higgs Centre for Theoretical Physics, 
School of Physics and Astronomy, \\
The University of Edinburgh, Edinburgh EH9 3FD, Scotland, UK}

\emailAdd{samuel.abreu@uclouvain.be}
\emailAdd{britto@maths.tcd.ie}
\emailAdd{claude.duhr@cern.ch}
\emailAdd{einan.gardi@ed.ac.uk}
\emailAdd{james.matthew@ed.ac.uk}

\abstract{
It is well known that Feynman integrals in dimensional regularization often
evaluate to functions of hypergeometric type. 
Inspired by a recent proposal for a coaction on one-loop Feynman integrals 
in dimensional regularization, we use 
intersection numbers and twisted homology theory to define a coaction 
on certain hypergeometric functions. The functions we consider admit an 
integral representation where both the integrand and the contour of integration 
are associated with positive geometries. As in dimensionally-regularized Feynman 
integrals, endpoint singularities are regularized by means of exponents controlled
by a small parameter $\eps$. We show that the coaction defined on this class of 
integral is consistent, upon expansion in~$\eps$, with the well-known coaction 
on multiple polylogarithms. We illustrate the validity of our construction by 
explicitly determining the coaction on various types of hypergeometric ${}_{p+1}F_p$ and Appell functions.
}


\begin{document}
\maketitle





\section{Introduction}

Feynman integrals are a cornerstone of perturbative quantum field
theory and they are ubiquitous when evaluating higher orders in the 
perturbative series.
As such, having efficient tools for their evaluation and a thorough 
understanding of their mathematical properties is of great importance. 
For this reason, Feynman integrals and their mathematical structure are 
an active field of study both in mathematics and physics.

It follows from unitarity that Feynman integrals 
must be transcendental functions, as they must have 
nonvanishing discontinuities stemming from logarithmic branch cuts. 
In integer dimensions, the class of transcendental functions that can arise is further constrained to 
be \emph{periods}~\cite{BelkaleBrosnan,Bogner:2007mn}, which are
integrals of algebraic functions over domains defined by inequalities between algebraic functions~\cite{periods}.
Periods are interesting objects in their own right in mathematics, and it is 
known that they can be equipped with a lot of algebraic structure.
Of particular interest for this paper will be the so-called coaction, see for example ref.~\cite{Brown:coaction}. 
It was shown in ref.~\cite{Brown:2015fyf} that, quite generically, this 
algebraic structure and the coaction are inherited by the Feynman integrals
themselves. 
Understanding these structures in detail may open the way to a novel 
understanding of perturbative quantum field theory. 
For first applications in a physics context, see, e.g., 
refs.~\cite{Panzer:2016snt,Schnetz:2017bko,Caron-Huot:2019bsq}.

While the algebraic structures on Feynman integrals introduced in 
ref.~\cite{Brown:2015fyf} apply in a broad range of cases, 
they ignore a key aspect of Feynman 
integrals arising in physically-relevant perturbative quantum 
field theories, namely the unavoidable fact that these involve
divergent integrals in four space-time dimensions.
In order to make sense of divergent integrals one needs to 
introduce a suitable regulator.
A consistent framework for regularizing the divergences is provided by
dimensional regularization, where the integral is computed
in $D=4-2\epsilon$ dimensions. The Feynman integral is then
a meromorphic function of $\epsilon$, as can be seen, for instance,
from the so-called Feynman-parametric representation. 
Singularities may occur at rational values of $\epsilon$, and those at 
$\epsilon=0$ encode the divergences of the integrals in the 
four-dimensional limit.
The route leading from divergent Feynman integrals to finite physical quantities 
is rather involved: ultraviolet singularities are eliminated in the process of 
renormalization, while infrared ones
cancel in infrared-finite observables. 
As a consequence, dimensionally-regularized Feynman integrals are 
an integral part of most modern approaches to compute higher-order corrections
in perturbation theory.

As functions of~$\eps$, Feynman integrals are not 
periods. Instead, it is the coefficients in their 
Laurent expansion in $\eps$ that are 
periods~\cite{Bogner:2007mn}. Algebraic structures, such as the coaction,
that have been defined for periods do not directly 
extend to dimensionally-regularized integrals,
and one can only apply the coaction order by order in $\eps$.
However, if one believes that the coaction is an intrinsic property
of the Feynman integrals themselves, one might expect that there should be a
way to extend it beyond the formalism developed for periods.
First steps in that direction were taken in \cite{Abreu:2017enx,Abreu:2017mtm}, where
we conjectured a formula for a coaction that maps
integrals into pairs of integrals obtained from a basis of integrands 
$\{\omega_i\}$  and integration contours $\{\gamma_j\}$ according to
\begin{equation}\label{eq:masterformula}
	\Delta\left(\int_\gamma\omega\right)=
	\sum_{ij}c_{ij}\int_\gamma\omega_i\otimes
	\int_{\gamma_j}\omega\,.
\end{equation}
More precisely, the $\{\omega_i\}$ are forms that generate the cohomology 
group associated with the integral on the left-hand side, and the $\{\gamma_j\}$ 
are cycles that generate the corresponding homology group. 
In refs.~\cite{Abreu:2017enx,Abreu:2017mtm} the matrix $c_{ij}$ 
was computed using an operation called semi-simple projection.
In the case of one-loop integrals, the first entry of the tensor in the coaction
was identified as the integral associated with  a contraction of the 
original Feynman graph, and the 
second entry as a cut of the Feynman integral \cite{Froissart,Abreu:2017ptx}, leading
to an elegant diagrammatic representation for the coaction.
One-loop Feynman integrals have the special property
that the periods that appear in their Laurent expansion in $\epsilon$ are all
multiple polylogarithms (MPLs), and, remarkably, the coaction obtained from eq.~\eqref{eq:masterformula} was shown to be consistent with the expansion in 
$\epsilon$. More precisely, it was observed that
if both sides of the equality in eq.~\eqref{eq:masterformula} 
are expanded in $\epsilon$, then the coaction in eq.~\eqref{eq:masterformula} 
reproduces the coaction on MPLs.

It is well known that hypergeometric-type integrals
(see, e.g.,~ref.~\cite{Erdelyi_Transcendental}) appear when
evaluating Feynman integrals in dimensional regularization.
These functions depend on a set of parameters and
a set of variables. For instance, the well-known Gauss hypergeometric function $_2F_1(\alpha,\beta;\gamma;x)$
depends on the parameters $\a,\b,\c$ and the variable $x$; in the Euler-type integral representation we will be using, given in eq.~(\ref{2F1}) below, the former parametrize the exponents governing the powers of polynomial functions of the latter.  In the context of Feynman integrals,
the parameters are linear in the dimensional regulator~$\epsilon$,
and the variables depend on the kinematics of the Feynman diagram. 
The specific type of hypergeometric function also depends on the diagram considered. Let us recall a few examples from the literature: for one-loop integrals in general kinematics one finds the ${}_2F_1$ function in two-point functions (see e.g.~the appendix
of~ref.~\cite{Abreu:2017mtm}), the so-called Appell $F_1$ function in three-point
functions and the Lauricella-Saran $F_S$ or $F_N$ functions \cite{saran1955}
in four-point functions (see e.g. ref.~\cite{Fleischer:2003rm,Davydychev:2017bbl}). Higher-point integrals have not been computed explicitly in general kinematics as a function of $\epsilon$, but for example the massless pentagon evaluates to Appell $F_3$ functions \cite{Kniehl:2010aj}.
Beyond one loop, one also finds that Feynman integrals
evaluate to similar classes of hypergeometric functions. For instance, two-loop sunrise integrals evaluate to either $_2F_1$, Appell $F_2$ or Appell $F_4$ functions, 
depending on the configuration of the masses of the propagators,
see e.g.~ref.~\cite{Tarasov:2006nk}. 
These examples illustrate the fact that hypergeometric-type integrals
are the functions we must understand when studying Feynman integrals
in dimensional regularization.

Building on the results
of refs.~\cite{Abreu:2017enx,Abreu:2017mtm}, where
one-loop Feynman integrals were observed to admit a diagrammatic coaction, valid to all orders in the dimensional regulator, 
it is natural to expect that one could define a coaction that acts on the relevant functions, 
independently of whether they appear in a Feynman integral.
By imposing restrictions on the form of the parameters, we will
focus on cases where the hypergeometric functions expand to MPLs,
so that we can explicitly verify that the coaction
we obtain from \eqref{eq:masterformula} reduces to the coaction
on MPLs upon expansion in $\epsilon$.
Constructing such a coaction is the main goal of this paper. It is 
important both for the study of hypergeometric-type integrals 
and in view of the possible 
extension of the diagrammatic coaction beyond one loop. 

The starting point for constructing the coaction in all 
cases of hypergeometric functions we will address
is their integral representation. 
We find the concept
of positive geometries~\cite{Arkani-Hamed:2017tmz} very
useful to study these integrals, because it allows us to find convenient bases of for the 
homology and cohomology groups of the corresponding integral
representations. With these bases, we can directly use the general
formula in eq.~\eqref{eq:masterformula} to obtain a coaction on the corresponding integrals.

In this paper, we also introduce a new feature in the construction
of the coaction in eq.~\eqref{eq:masterformula}. The matrix $c_{ij}$
is constructed by computing the matrix of intersection numbers between the
generators of the cohomology group $\omega_i$ and 
a set of forms $\Omega(\gamma_i)$ which,
under certain conditions, can be constructed in
a canonical way from the contours $\gamma_i$.
This is always possible to do for integrals defined by 
positive geometries,
which includes all examples we will address in this paper and for which
there is an explicit way of computing the forms $\Omega(\gamma_i)$.
Because we consider functions prior to expansion in $\epsilon$, the
integrands are themselves multi-valued functions. This
implies that we cannot use standard (co)homology theory to construct
our bases of forms and cycles. 
Instead, we must use the framework of `twisted (co)homology' \cite{AomotoKita}
which is well known in the mathematics literature. Recently,
these tools have been applied in several areas of theoretical physics such
as in string theory \cite{Mizera:2017rqa,Mizera:2019gea} or in the study of the 
integration-by-parts relations satisfied by Feynman 
integrals~\cite{Mastrolia:2018uzb,Frellesvig:2019kgj,Frellesvig:2019uqt}. 
Compared with the construction of ref.~\cite{Abreu:2017enx,Abreu:2017mtm},
where the normalization was based on a semi-simple projection, 
the approach we present here has the advantage of treating the
generators of the homology and cohomology groups on the same footing,
in the sense that the matrix $c_{ij}$ can be viewed as a change of basis
of the generators of either group.

We would like to mention that, following discussions about the content of refs.~\cite{Abreu:2017enx,Abreu:2017mtm,Abreu:2018nzy,Abreu:2018sat,ethtalk}
and parts of the content of the present paper, the authors of ref.~\cite{brown2019lauricella}
have initiated a rigorous mathematical treatment of the concepts  presented in this paper, specialized to a class of 
one-dimensional integrals representing Lauricella $F_D$ functions.

The paper is organized as follows. In section \ref{sec:mpsResum} we
summarize the coaction on MPLs and give first examples of a coaction acting
on an unexpanded function of $\epsilon$, obtained by resumming the Laurent series
of the integrals. In section~\ref{sec:construction_of_the_coaction} we discuss
positive geometries in order to define the type of integrals that we will consider 
in this paper, and we very briefly introduce the elements of
twisted (co)homology theory that will be relevant for this paper. 
Section~\ref{sec:coactionOnIntegrals} contains the main result of the paper,
namely the formula for a coaction that acts on unexpanded $\epsilon$-dependent
integrals, while being consistent with the Laurent expansion in~$\eps$. The remaining
sections contain examples in a variety of hypergeometric-type integrals.
We first discuss in detail Gauss' hypergeometric function $_2F_1$ in
section~\ref{sec:2F1}. In section~\ref{sec:one-dimensional} we discuss
a larger class of one-dimensional integrals depending on several variables, namely the Lauricella $F_D$ functions.
In section~\ref{sec:two-dimensional} we discuss two-dimensional integrals,
focusing on the Appell functions $F_1$, $F_2$, $F_3$ and $F_4$. Finally,
in section~\ref{sec:pp1fp} we discuss generic $_{p+1}F_p$ hypergeometric
functions, which are given by a $p$-dimensional integral.
In section~\ref{sec:conclusions} we summarize and discuss our results.


\section{The coaction on MPLs and resummation
of the $\epsilon$ expansion}
\label{sec:mpsResum}

In this section we give a brief review of MPLs and their coaction to establish
our notation. We then discuss some simple examples of Feynman integrals
in dimensional regularization and hypergeometric functions
where one can `resum' the Laurent series around $\epsilon=0$ to obtain a coaction
that is consistent with the Laurent expansion.

\subsection{The coaction on MPLs}

When considered order-by-order in dimensional regularization,
many multi-loop Feynman integrals can be evaluated in terms of MPLs, 
defined by the iterated integrals~\cite{2001math......3059G}
\beq\label{eq:MPL_def}
G(a_1,\ldots,a_n;z) = \int_0^z\frac{dt}{t-a_1}\,G(a_2,\ldots,a_n;t)\,,
\eeq 
where the $a_i$ and $z$ are (algebraic) complex numbers. In the case where all $a_i=0$, the integral in eq.~\eqref{eq:MPL_def} diverges,
and instead we define 
\beq\label{eq:G0_log}
G(\vec{0}_n;z)= \frac{1}{n!}\log^nz\,,\qquad \vec{0}_n=(\underbrace{0,\ldots,0}_{n\textrm{ times}})\,.
\eeq
MPLs are well studied in both the mathematics and physics literature (see, e.g., ref.~\cite{Duhr:2014woa} and references therein). 
In particular, they can be endowed with a coaction~\cite{2001math......3059G,2002math......8144G,B:MTMZ}, which we denote here by $\DeltaMPL$. Roughly speaking, the coaction associates to an MPL a linear combination of
tensor products of these functions. For example, the coactions of the logarithm in eq.~\eqref{eq:G0_log} or of the classical polylogarithm $\textrm{Li}_n(z) = -G(\vec{0}_{n-1},1;z)$ are given by
\beq\bsp\label{eq:Delta_MPL_examples}
\DeltaMPL(\log^n z)&\, = \sum_{k=0}^n\binom{n}{k}\log^{n-k} z\otimes \log^k z\,,\\
\DeltaMPL(\textrm{Li}_n(z))&\, = 1\otimes\textrm{Li}_n(z) + \sum_{k=0}^{n-1}\textrm{Li}_{n-k}(z)\otimes\frac{\log^kz}{k!}\,.
\esp\eeq
The formula for the coaction of a general MPL is more involved, and we refer to the literature for a discussion of the general case~\cite{2001math......3059G,2002math......8144G,B:MTMZ}. 

An important feature of the coaction on MPLs is that the second factor of each tensor is interpreted modulo its branch cuts. Since all discontinuities of MPLs are proportional to powers of $i\pi$, this is equivalent to setting to zero all factors of $i\pi$ in the second factor of each tensor in the coaction. The coaction  also operates nontrivially on transcendental constants obtained by specialising the arguments of the MPLs to some special values. In particular, at $z=1$ the classical polylogarithms reduce to zeta values, $\zeta_n = \textrm{Li}_n(1)$. For $n$ odd, the coaction of $\zeta_n$ is simply obtained by specialising eq.~\eqref{eq:Delta_MPL_examples} to $z=1$,
\beq\label{eq:zetaOdd}
\DeltaMPL(\zeta_n) = \zeta_n\otimes 1+ 1\otimes \zeta_n\,,\qquad n \textrm{ odd}\,.
\eeq
For $n$ even, the situation is more subtle, and we have to define~\cite{B:MTMZ,2011arXiv1102.1310B}
\beq\label{eq:zetaEven}
\DeltaMPL(\zeta_n) = \zeta_n\otimes 1\,,\qquad n \textrm{ even}\,,
\eeq
and more generally
\beq
\DeltaMPL(i\pi) = i\pi\otimes 1\,.
\eeq
These definitions are consistent with the fact that we have to work modulo factors of $i\pi$ in the second factor.

\subsection{Resummation of the $\epsilon$ expansion}

When working in dimensional regularization, MPLs appear as the Laurent coefficients in the $\epsilon$ expansion, and we can only consider the coaction order by order in the expansion. A natural question to ask is if one can `resum' the Laurent series after acting with $\DeltaMPL$ on its coefficients. To illustrate this point, let us consider the simplest  Feynman integral, namely the tadpole integral with one massive propagator in $D=2-2\epsilon$ dimensions,
\beq\bsp
T(m^2,\eps) &\,= \frac{e^{\gamma_E\eps}}{i\pi^{D/2}}\int\frac{d^Dk}{k-m^2} = \frac{e^{\gamma_E\eps}\,\Gamma(1+\eps)}{\eps(1-\eps)}\,m^{-2\eps}\\
&\,=\frac{1}{\eps} + 1-\log m^2 +\eps\left(\frac{1}{2}\log^2m^2 - \log m^2 +1+\frac{\pi^2}{12}\right)+\ord(\eps^2)\,,
\esp\eeq
where $\gamma_E=-\Gamma'(1)$ denotes the Euler-Mascheroni constant. We can act with $\DeltaMPL$ order by order in the expansion, and it is straightforward to check that at each order the resulting formula is consistent with a `resummed' coaction, 
\beq
\Delta(T(m^2,\eps)) = \frac{1}{\eps(1-\eps)}\,\left[e^{\gamma_E\eps}\,\Gamma(1+\eps)\,m^{-2\eps}\right]\otimes \left[e^{\gamma_E\eps}\,\Gamma(1+\eps)\,m^{-2\eps}\right]\,.
\eeq
In fact, it is easy to prove the previous formula by using the fact that $\Delta_{}(a\cdot b)=\Delta_{}(a)\cdot \Delta_{}(b)$ as well as
\beq\bsp\label{eq:Delta_eps_example_1}
\Delta\left(\frac{1}{\eps(1-\eps)}\right) &\,= \frac{1}{\eps(1-\eps)}\,1\otimes 1\,,\\
\Delta(m^{-2\eps}) &\,= m^{-2\eps}\otimes m^{-2\eps}\,,\\
\Delta\left[e^{\gamma_E\eps}\,\Gamma(1+\eps)\right] &\,= \left[e^{\gamma_E\eps}\,\Gamma(1+\eps)\right]\otimes \left[e^{\gamma_E\eps}\,\Gamma(1+\eps)\right]\,.
\esp\eeq
These formulas are obtained by expanding the argument of $\Delta$ in $\eps$ and using the linearity of the coaction.
For instance,
\beq\bsp
\Delta\left(\frac{1}{\eps(1-\eps)}\right) &\,= \sum_{k=0}^{\infty}\eps^{k-1}\,\DeltaMPL(1) = \sum_{k=0}^{\infty}\eps^{k-1}\,(1\otimes1) = \frac{1}{\eps(1-\eps)}\,1\otimes 1\,,\\
\Delta(m^{-2\eps}) &\,= \sum_{k=0}^\infty\frac{(-\eps)^k}{k!}\,\DeltaMPL(\log^km^2) = \sum_{k=0}^\infty\sum_{l=0}^k\frac{(-\eps)^k}{k!}\,\binom{k}{l}\,\log^{k-l}m^2\otimes \log^lm^2\\
&\, = \sum_{k,l=0}^\infty \frac{(-\eps)^{k+l}}{k!\,l!}\,\log^{k}m^2\otimes \log^lm^2 = m^{-2\eps}\otimes m^{-2\eps}\,,
\esp\eeq
where we have used eq.~\eqref{eq:Delta_MPL_examples} in the second line.
The formula for the coaction of the gamma function can be proven in the same way using the well-known formula
\beq
e^{\gamma_E\eps}\,\Gamma(1+\eps) = \exp\sum_{k\ge2}\frac{(-\eps)^k\,\zeta_k}{k}\,,
\eeq
together with eqs.~\eqref{eq:zetaOdd} and \eqref{eq:zetaEven}.

This example shows that in the case of the tadpole integral, it is possible to `resum' the $\eps$-expansion 
of the coaction to obtain a version of the coaction valid to all orders in~$\eps$.
	We note that, for these simple functions, this procedure can in fact be given a rigorous mathematical
	grounding in terms of the motivic coaction on multiple-zeta values and on the
	logarithm \cite{brownLetter}.
In more general 
cases, however, the functional dependence of the integral on the kinematic variables before expansion 
in $\eps$ is much more complicated and often involves functions of hypergeometric type. 

The simplest nontrivial hypergeometric function is Gauss' ${}_2F_1$ function.
This function admits an Euler-type integral representation of the form
\begin{equation}
\label{2F1}
 {}_2F_1(\a,\b;\c;x)=\frac{\Gamma(\c)}{\Gamma(\a)\Gamma(\c-\a)}
 \int_0^1du\,u^{\a-1}(1-u)^{\c-\a-1}(1-ux)^{-\b}\,
\end{equation}
provided that the integral converges. Here we focus on
a subset of cases, namely those where $\a,\b,\c$ have the form $m+a\eps$, where
$m\in\mathbb{Z}$. Under these restrictions, its
Laurent expansion in $\eps$ involves MPLs as coefficients (which can be computed 
in an algorithmic way, cf.~ref.~\cite{Moch:2001zr,Moch:2005uc,Weinzierl:2002hv,Huber:2005yg,Kalmykov:2006pu}), and we can act with 
$\DeltaMPL$  on the coefficients order by order in the expansion. It is  
tantalizing to speculate whether it is possible to `resum' the Laurent 
expansion after acting with the coaction. The central proposal of this paper is that, remarkably, this is indeed possible. Before we dive into the 
mathematical formalism in the next section, 
let us illustrate this result on a special case of a ${}_2F_1$ function for
which the Laurent coefficients can be written in closed form~\cite{Bern:1999ry},
\beq\label{eq:2F1Li}
{}_2F_1(-\eps,1;1-\eps;x) = 1-\sum_{n=1}^{\infty}\eps^n\,\textrm{Li}_n(x) 
= 1-F(\eps,x)\,.
\eeq
Using eq.~\eqref{eq:Delta_MPL_examples} for the coaction of the polylogarithms, we easily obtain
\beq\label{eq:Delta_eps_example_2}
\Delta\left[F(\eps,x)\right] =  1\otimes F(\eps,x) + F(\eps,x)\otimes x^{\eps}\,,
\eeq
where we the coaction $\Delta$ acts order by order in the $\eps$ expansion.

In the remainder of this paper we argue that formulas like eq.~\eqref{eq:Delta_eps_example_1} and~\eqref{eq:Delta_eps_example_2} are not the exception but rather the rule, at least for very large classes of hypergeometric functions whose Laurent expansion can be expressed in terms of MPLs. More precisely, we argue that, for certain classes of functions, we can define a coaction $\Delta_\eps$ valid to all orders in the~$\eps$ expansion. In our examples, the form of this coaction is obtained by replacing $\Delta_{}$ by $\Delta_\eps$ everywhere in~eq.~\eqref{eq:Delta_eps_example_1} and~\eqref{eq:Delta_eps_example_2}. The Laurent expansion around $\eps=0$ then `intertwines' the two coactions, i.e., if $L_{\eps}$ denotes the operator which assigns to a function its Laurent expansion, we have
\beq\label{eq:coactionLaurentMot}
(L_{\eps}\otimes L_{\eps})\Delta_{\eps} = \DeltaMPL L_{\eps}\,,
\eeq
where in the right-hand side $\DeltaMPL$ acts order-by-order in the Laurent expansion.


\section{Positive geometries and canonical forms}
\label{sec:construction_of_the_coaction}

In this section we define the class of integrals that we will address in this
paper, introduce some of the mathematical background and
establish our notation. We will use the example of Euler's beta function 
to illustrate the different objects we introduce.

\subsection{A class of integrals}
\label{sec:class_of_integrals}

Let us consider an integral obtained by integrating a differential form $\omega$ over some domain $\gamma$.
We start by characterizing the classes of integrands that will be of relevance in this paper.
Our integrand $\omega$ depends on
$n$ integration variables $u_i$, $i=1,\ldots,n$ as well as external
variables $x_j$, $j=1,\ldots,m$ which we do not write explicitly.
More precisely, $\omega$ is an $n$-form
\begin{equation}\label{eq:factorizedIntegrand}
	\omega=d\bold{u}\prod_{I} P_I(\bold{u})^{\alpha_I}\,,
\end{equation}
with $d\bold{u}=du_1\wedge\ldots\wedge du_n$,
where the $P_I$ are polynomials in the $u_i$ and $x_j$, 
and with $\alpha_I\in\mathbb{C}$. 
We assume the $P_I(\bold{u})$ to be irreducible over
rational functions of the $x_j$.
Furthermore, we assume that the exponents take the form
$\alpha_I=n_I+a_I\epsilon$, with $n_I\in\mathbb{Z}$,
$a_I\epsilon\in\mathbb{C}^*$, $\sum_I a_I \neq 0$, and where $\epsilon$ can be taken to be infinitesimally small. 
Finally, we define the decomposition
$\omega=\Phi\varphi$ where
\begin{equation}
\label{eq:Phiphi}
\Phi = \prod_{I} P_I({\bf u})^{a_I \eps}\quad \textrm{and} \quad
\varphi = d{\bf u} \prod_{I} P_I({\bf u})^{n_I}\,.
\end{equation}
We will always further restrict the form of these polynomials such that the coefficients
in the Laurent expansion of the integrals $\int_{\gamma}\omega$ in $\eps$ only involve MPLs. We will be more specific about the form of the polynomials in subsequent sections.

The integration contour $\gamma$
is a $n$-dimensional cycle in
\begin{equation}\label{eq:Xdef}
	X(\mathbb{C})=\cproj{n}\setminus\bigcup\limits_{I} \{P_I(\bold{u})=0\}\,,
\end{equation}
where $\cproj{n}$ is the $n$-dimensional complex projective space.\footnote{Strictly speaking, $\{P_I(\bold{u})=P_I(u_1,\ldots,u_n)=0\}$ is an affine variety in $\mathbb{C}^n$. We use the same notation for the affine variety and its lift to projective space.} In other words, $\gamma$ is a domain with boundary contained in the union of the varieties defined by $P_I({\bf u})=0$. Since $a_I\neq0$, then 
$\Phi$ vanishes on the boundary of $\gamma$, at least for some ranges of values of $\epsilon$, and thus for all values by analytic continuation. As a consequence, there are no boundary
contributions when performing integration by parts.

The natural mathematical framework to discuss such integrals is that
of twisted homology and cohomology 
\cite{AomotoKita}---see also 
refs.~\cite{Mizera:2017rqa,Mastrolia:2018uzb,
Frellesvig:2019uqt,Frellesvig:2019kgj}. 
We define the \emph{twist} $d\log\Phi$ (we will often call $\Phi$ the twist)
and consider the covariant differential
\begin{equation}
	\nabla_\Phi=d+d\log\Phi\wedge\,.
\end{equation}
We then have $d(\Phi\xi) = \Phi\,\nabla_{\Phi}\xi$, where $\xi$ can be any smooth differential form. Stokes' theorem implies that
for an arbitrary smooth $(n-1)$-form $\xi$ we have
\begin{equation}
	\int_\gamma\Phi\varphi=
	\int_\gamma\Phi(\varphi+\nabla_\Phi\xi)\,.
\end{equation}
As a consequence, the integrand is only defined up to adding a total covariant derivative, and 
 we are therefore interested in the (twisted) cohomology groups
\begin{equation}\label{eq:H_dR_def}
	H^n(X,\nabla_\Phi)=
	\{\varphi|\nabla_\Phi\varphi=0\}
	/
	\{\nabla_\Phi\xi\}\,.
\end{equation}
Similarly, we can construct twisted homology groups by considering
twisted cycles in $X(\mathbb{C})$~\cite{AomotoKita}. The precise definition of the twisted cycles is not important for the rest of the paper. Here it suffices to say that they can be thought of as ordinary cycles $\gamma$, together with information on the Riemann sheet (with respect to the multi-valued function $\Phi$) on which this cycle is considered.

The cohomology groups in eq.~\eqref{eq:H_dR_def} will play a prominent role in 
this paper. We will be particularly interested in determining explicit bases for 
the cohomology group associated to the integral we want to consider,
which is in general an extremely difficult problem to solve.
In some examples, however, it is easy to determine at least the dimension of
these groups.
One can start by counting the critical points of the function $\Phi$, i.e.~the 
number of independent solutions to the equation
\begin{equation}
	d\log\Phi=0\,,
\end{equation}
which gives an upper bound for the dimensionality. 
This upper bound is saturated under certain conditions 
outlined for example in ref.~\cite{AomotoKita}, 
see also refs.~\cite{Lee:2013hzt,Bitoun:2018afx,Frellesvig:2019kgj}.
In particular, the bound is saturated in each of the 
examples studied in this paper.

\subsection{Integrals and positive geometries}
\label{sec:positiveGeom}

As already mentioned, it can be very difficult to construct explicit bases for the cohomology groups. Here, however, we are not interested in the most general case, but we restrict our attention to those cases where they are generated by (wedge products of) $d\log$-forms, i.e., by differential forms with logarithmic singularities along the varieties $\{P_I({\bf u})=0\}$.

A particularly convenient such geometric setting 
is that of $\emph{positive geometries}$ \cite{Arkani-Hamed:2017tmz},
whose definition we briefly recall in this section.
Let $Y(\mathbb{C})$ be an irreducible complex projective variety of dimension
$n$, corresponding to the solution in $\cproj{n}$ of 
homogeneous polynomial equations with rational coefficients. We assume that $Y(\mathbb{C})$ has no nonzero
holomorphic $n$-forms.
We denote by $\rproj{n}$ the $n$-dimensional real projective space, 
and by $Y(\mathbb{R})$ the real part of $Y(\mathbb{C})$, i.e.,
the solution of the same polynomial
equations in $\rproj{n}$.
For concreteness, we will always be working in an affine chart of projective space, with affine coordinates ${\bf u} = (u_1,\ldots,u_n)$.
In this chart, the surfaces are described by the polynomial equations $P_I(\bold{u})=0$, which carve out
$n$-dimensional cells $\Gamma_j$ in $Y(\mathbb{R})$. 
In the following we will always be interested in the case
where $Y(\mathbb{C})=\cproj{n}$.
A positive geometry is a pair $(\cproj{n},\Gamma_j)$
together with a differential
form $\Omega(\cproj{n},\Gamma_j)$, called the \emph{canonical form}, with simple poles on the boundary
of $\Gamma_j$. The form $\Omega(\cproj{n},\Gamma_j)$ is unique 
(up to normalization) because we assume that there are no nonzero
holomorphic $n$-forms.
We also require that all boundary components are themselves positive geometries whose canonical form is given 
by\footnote{To be more precise, recall that we have assumed the $P_I(\bold{u})$ to be
	irreducible polynomials. Then the boundary component $\partial_I\Gamma_j$
	of $\Gamma_j$ is the part of the boundary of $\Gamma_j$ that lies
	in the variety $\{P_I(\bold u)=0\}$. The
$\partial_I\Gamma_j$  are themselves positive geometries
$(\cproj{n-1},\partial_I\Gamma_j)$ with associated canonical forms.}
\begin{equation}\label{eq:recursiveCanForms}
\Omega(\cproj{n-1},\partial_I\Gamma_j)=\res_{\partial_I\Gamma_j}\Omega(\cproj{n},\Gamma_j)\,.
\end{equation}
We refer the reader to ref.~\cite{Arkani-Hamed:2017tmz} for a more precise definition.

For $n=0$, the $\Gamma_j$ are points and $\Omega(\cproj{0},\Gamma_j)=\pm1$ 
for any $j$. For $n>0$, we can choose the forms $\Omega(\cproj{n},\Gamma_j)$ 
to be $d\log$-forms.
In the following,
we will refer to the cells $\Gamma_j$ themselves
as positive geometries, and we will denote their canonical forms simply by $\Omega(\Gamma_j)$.
We note that the map $\Omega$, which associates to a positive geometry $\Gamma_j$ its canonical form, is closely related to the map $c_0$ that has appeared in ref.~\cite{Brown:2018omk}.

For the integrals considered in this paper, 
the $\Gamma_j$ and the associated canonical forms $\Omega(\Gamma_j)$ are natural
candidates for the generators of the (co)homology groups associated to the geometry underlying the  integral under consideration. In other words, we will be considering integrals of the form 
\begin{equation}
	\int_{\Gamma_{j'}}\Phi\,\Omega(\Gamma_j)\,.
\end{equation}
Integrals of this type, where we integrate the canonical form of the positive geometry $\Gamma_j$ over some other positive geometry $\Gamma_{j'}$ were called \emph{canonical integrals} in ref.~\cite{Arkani-Hamed:2017tmz}.
Here we slightly generalize this notion and consider canonical integrals with a twist $d\log\Phi$ \cite{Mizera:2017rqa,Mastrolia:2018uzb}.
For generic $a_I$ in $\Phi$, the integral above is well-defined even if $j=j'$,
as long as it converges. The result can then be analytically continued
to any $a_I\in\mathbb{C}$ (up to poles).

{In general, not all canonical integrals are independent 
and it is convenient to find
bases $\vec\gamma\equiv(\gamma_1,\ldots,\gamma_k)$ 
of the associated homology group and
$\vec\varphi\equiv(\varphi_1,\ldots,\varphi_k)$ of the associated 
cohomology group. The elements of the cohomology group are equivalence 
classes of differential forms, and since we restrict ourselves to
canonical integrals, each class contains a $d\log$ representative. 
Throughout this paper we simply represent each class by this $d\log$ form,
keeping in mind that the elements of the cohomology groups are actually 
equivalence classes and not differential forms. Similarly, we will represent 
elements of the homology groups simply by the cycles $\Gamma_j$.}

As already stated at the end of the previous section, the dimension $k$ of the homology and cohomology groups can be determined by counting the critical
points of $\Phi$. 
Any cell $\Gamma_j$ can then be written as a linear combination of 
the basis elements $\gamma_i$, and similarly for the associated canonical 
forms and the basis $\varphi_i$.
It is known how to find such bases in the case where the $P_I(\bold{u})$
are either all linear (an ``arrangement of hyperplanes'') or all linear but one, where the latter can have degree up 
to $n$ \cite{Arkani-Hamed:2017tmz,Mastrolia:2018uzb}. 
In practice, this is sufficient for all the 
examples we will be interested in this paper.
For instance, the homology group of intersections of hyperplanes in general
position\footnote{When the hyperplanes are not in
general position there may be non-normal crossings, in which case
we need to perform a blow-up around the non-normal crossing
surface or, using physicists' nomenclature, use sector
decomposition to resolve the overlapping singularities.} 
is generated by \emph{bounded chambers} (the cells
that do not extend to infinity)~\cite{AomotoKita}.
A cycle $\Gamma$ defined by having its boundaries on hyperplanes 
is a positive geometry whose canonical form can be written down explicitly as follows.
If the hyperplanes at the boundaries are defined by the linear 
equations $P_1({\bf u}) = \ldots=P_{k}({\bf u}) = 0$, then the associated 
canonical form is uniquely determined up to a sign to be
\beq
\Omega(\Gamma) = d\log\frac{P_2({\bf u})}{P_1({\bf u})}\wedge\ldots 
\wedge d\log\frac{P_{k}({\bf u})}{P_{k-1}({\bf u})}\,.
\eeq

As an example, consider the case where $n=1$, i.e., there is a single integration
variable~$u$.
The solutions to $\prod_I P_I(u)=0$
are points $u_i$, which are a special case of hyperplanes in
 $\rproj{1}$. 
The bounded chambers are the 
 intervals between consecutive finite solutions, but 
any interval
$[u_i,u_j]$ can be seen as a positive geometry with the canonical form
\begin{equation}\label{eq:conventionsCan}
	\Omega([u_i,u_j])=d\log\frac{u-u_i}{u-u_j}\,.
\end{equation}
The boundary components of $[u_i,u_j]$ are the points
$u_i$ and $u_j$, which are 0-dimensional positive geometries
with the canonical forms 
\begin{equation*}
	\res_{u=u_i} \Omega([u_i,u_j])=
	\Omega(\{u_i\})=1\,,\qquad
	\res_{u=u_j} \Omega([u_i,u_j])=
	\Omega(\{u_j\})=-1\,,
\end{equation*}
in agreement with the recursive construction of eq.~\eqref{eq:recursiveCanForms}.
We can also consider cycles extending to infinity, which are also positive geometries whose canonical forms  are given by
\begin{equation}\label{eq:conventionsCanInfty}
	\Omega(\left[u_i,\infty\right))=d\log(u-u_i)\,.
\end{equation}

\paragraph{Example: Euler's beta function.} 

Consider the integral
\begin{equation}\label{eq:genBetaFunc}
\int_\gamma \omega  = \int_{\gamma} u^{\a_1}(1-u)^{\a_2} du\,.
\end{equation}
This integral contains as a special case
Euler's integral representation of the beta function,
\begin{equation}\label{eq:betaGenArg}
B(\a,\b) = \int_0^1 u^{\a-1}(1-u)^{\b-1} du\,,
\end{equation}
which converges for $\Re(\a)>0$ and $\Re(\b)>0$. 
The integrand $\omega$ has two linear factors,
$P_1(u) = u$ and $P_2(u)=1-u$,
raised to the  exponents
$\a_1=n_1+a_1\eps$ and,  $\a_2=n_2+a_2\eps.$
The multivalued function $\Phi$ is 
$
\Phi = u^{a_1\eps}(1-u)^{a_2\eps}.
$

There is a single bounded chamber, $[0,1]$, defined by the polynomials 
$P_1(u)$ and $P_2(u)$.
This is consistent  with the
fact that there is a single solution to $d\ln\Phi=0$, i.e,
$\Phi$ has a single critical point. The associated canonical form is
\begin{equation}\label{eq:basisCohomBeta}
	\Omega([0,1])=d\log\frac{u}{u-1}\,.
\end{equation}
Note that there are other positive geometries we could have considered
(another example is $[1,\infty]$), but since $P_1(u)$ and $P_2(u)$
are linear in $u$ it is sufficient to consider the bounded chamber.
Furthermore, through integration-by-parts relations, 
the one-form in eq.~\eqref{eq:basisCohomBeta} gives 
a basis of the first cohomology group of eq.~\eqref{eq:genBetaFunc}. In other
words, any integral of the type \eqref{eq:genBetaFunc} is a multiple of
\begin{equation}
	\int_{0}^1du\, u^{a_1\epsilon}(1-u)^{a_2\epsilon}
	\left(\frac{1}{u}-\frac{1}{u-1}\right)\,.
\end{equation}
This is straightforward to check by explicit calculation, using well-known identities among gamma functions.

\subsection{Pairings of cycles and forms}

Having discussed how to construct the twisted homology
and cohomology groups associated with a canonical integral, 
we now discuss how to pair elements of these groups through complex-valued bilinear maps.

We denote by $\vec\gamma$ a basis of the homology group
and by $\vec\varphi$ a basis of the cohomology group. The most
natural pairing  is to associate
cycles $\gamma_k\in\vec\gamma$ with forms
$\varphi_l\in\vec\varphi$ to compute the so-called
\emph{period matrix}
\begin{equation}\label{eq:periodMat}
	P_{kl}\!\left(\vec\gamma;\vec\varphi;\Phi\right)
	=\int_{\gamma_k}\Phi\varphi_l\,,
\end{equation}
where each row corresponds to a cycle and each column corresponds to a form.
The matrix $P$ is a square matrix whose dimension is given by the
dimension of the (co)homology group, which we recall
can be determined by counting the critical points of~$\Phi$.
For a given twist $\Phi$, cycle $\gamma$ and integrand $\omega=\Phi\varphi$, and assuming that
$\gamma$ is in the associated homology group
and $\varphi$ is in the associated cohomology group,
any integral $\int_\gamma\omega$ can
be written as a linear combination of the elements of the period
matrix,
\begin{align}\label{eq:genericAsPeriods}
\begin{split}
	\int_\gamma\omega
=&	\sum_{k,l}c_{kl}\,P_{kl}\!\left(\vec\gamma;\vec\varphi;\Phi\right)\,.
\end{split}
\end{align}
The algebraic properties of any
integral of this type can then be studied
from the entries of the period matrix.

A less obvious pairing we can construct is between
two differential forms. Let us assume we have
two bases $\vec\varphi$ and $\vec\psi$, not necessarily distinct, of the same twisted cohomology group.  We can then
compute \emph{intersection numbers} 
$\langle \varphi_i,\psi_j\rangle_{\Phi}$ between these forms.
To be more precise, 
we must first construct a dual twisted cohomology group, which
is also generated by $\vec\psi$ but for which
the covariant differential is $\nabla_{\Phi^{-1}}$. In our
case, this corresponds to taking $\epsilon\to-\epsilon$
in $\Phi$. We can then pair generators 
$\langle\varphi_i|$ of the cohomology with elements
$|\psi_j\rangle$ of the dual cohomology \cite{AomotoKita}
(see also ref.~\cite{Mizera:2017rqa}),
\begin{equation}\label{eq:intNumbDef}
	\langle \varphi_i,\psi_j\rangle_\Phi
	=\frac{1}{(2\pi i)^2}\int_{X(\mathbb{C})}
	\iota_\Phi(\varphi_i)\wedge\psi_j\,,
\end{equation}
with $X(\mathbb{C})$ as defined in eq.~\eqref{eq:Xdef} and
$\iota_\Phi$ the map that associates to a form $\varphi_i$
a form $\iota_\Phi(\varphi_i)$ in the same cohomology class
but with compact support so that the integral is well 
defined~\cite{Matsumoto1998,Mizera:2017rqa}. 
Intersection
numbers can then be arranged in the matrix
\begin{equation}\label{eq:C_matrix_def}
	C_{kl}\big(\vec\varphi;\vec\psi;\Phi\big)
	=\langle \varphi_k,\psi_l\rangle_\Phi\,,
\end{equation} 
which has the same dimensions as the period matrix $P$.

The matrix of intersection numbers $C\big(\vec\varphi;\vec\psi;\Phi\big)$  in eq.~\eqref{eq:C_matrix_def} will play a very important role 
in the construction of our coaction. 
Therefore we need efficient ways of computing intersection numbers. 
The definition of the intersection numbers in eq.~\eqref{eq:intNumbDef} 
is not the most convenient for practical calculations,
so we now discuss three alternative ways to compute them in the cases we are interested in.
In the case where $n=1$ and the $\varphi_i$ and $\psi_j$ 
are $d\log$-forms, which covers several of the examples of this paper,
a more explicit way to compute the intersection numbers is
\cite{Mizera:2017rqa,Mastrolia:2018uzb}
\begin{equation}
	\label{eq:intRes}
	\vev{\varphi_i,\psi_j}_\Phi
	=\sum_{u_p\in \mathcal{P}(\Phi)}
	\frac{\res_{u=u_p}\varphi_i\,\res_{u=u_p}\psi_j}
	{\res_{u=u_p}d\ln\Phi}\,,
\end{equation}
where 
$\mathcal{P}(\Phi)$ is the set of poles of $d\ln\Phi$.
This very explicit formula can be generalized to the case
where $n>1$ \cite{Frellesvig:2019uqt}.

Since the $\varphi_i$ and $\psi_j$ are not necessarily $d\log$-forms,
we can also use the alternative formulas proposed in ref.~\cite{Mizera:2017rqa}.
For $n=1$ and setting $u_1=u$, 
\bea
\label{eq:intnumbersell1}
\vev{\varphi_i,\psi_j}_\Phi = \sum_{u^*} \left(
\frac{\partial^2\log\Phi}{\partial{u}^2}
\right)^{-1}
\left.
\widehat\varphi_i \,\widehat\psi_j
\right|_{u=u^*}\,,
\eea
where the sum is over the critical points, i.e., the points $u^*$ satisfying 
$d\ln\Phi(u^*)=0$, and $\varphi_i=\widehat\varphi_i\,du$ and similarly
for $\psi_j$.
In the case $n=2$, with $(u_1,u_2)=(u,v)$, 
\begin{equation}\label{eq:intnumbersell2}
\vev{\varphi_i,\psi_j}_\Phi = \sum_{(u^*,v^*)} 
\det{}^{-1}
\left(
\begin{array}{cc}
 \frac{\partial^2\log\Phi}{\partial{u}^2} & 
 \frac{\partial^2\log\Phi}{\partial u \,\partial v} \\
 \frac{\partial^2\log\Phi}{\partial u \,\partial v} & 
 \frac{\partial^2\log\Phi}{\partial{v}^2}
\end{array}
\right)
\left.
\widehat\varphi_i \,\widehat\psi_j
\right|_{(u,v)=(u^*,v^*)}\,,
\end{equation}
where the sum extends over the critical points $(u^*,v^*)$ satisfying 
\[
\partial_u\log\Phi(u^*,v^*)=\partial_v\log\Phi(u^*,v^*)=0.\]

Finally, we can also use the period matrix to compute a certain
matrix of intersection numbers. Let
$P\!\left(\vec\gamma;\vec\varphi;\Phi\right)$ be the period matrix
constructed from the contours $\vec\gamma$ and $d\log$-forms
$\vec\varphi$ as defined above.
Then, the matrix $C\!\left(\Omega(\vec\gamma);\vec\varphi;\Phi\right)$ is
related to the period matrix $P\!\left(\vec\gamma;\vec\varphi;\Phi\right)$
through \cite{Mastrolia:2018uzb}
\begin{equation}\label{eq:periodIntersections}
	P\!\left(\vec\gamma;\vec\varphi;\Phi\right)=
	C\!\left(\Omega(\vec\gamma);\vec\varphi;\Phi\right)
	\Big(1+\mathcal{O}(\epsilon)\Big)\,,
\end{equation}
where we defined $\Omega(\vec\gamma) \equiv (\Omega(\gamma_1),\ldots,\Omega(\gamma_k))$.
We note that this relation is in agreement with the fact that 
the entries of $P\!\left(\vec\gamma;\vec\varphi;\Phi\right)$ are
multivalued functions while those of 
$C\!\left(\Omega(\vec\gamma);\vec\varphi;\Phi\right)$
are not. 
Indeed, given our choice of using a basis of $d\log$ forms, 
the leading order in the Laurent
expansion of $P\!\left(\vec\gamma;\vec\varphi;\Phi\right)$ is single-valued.

\paragraph{Example: Euler's beta function.} 

We return to the example of eq.~\eqref{eq:genBetaFunc}. In the previous
section we established that it is sufficient to study the beta function in
eq.~\eqref{eq:betaGenArg}. 
Building on the discussion on positive geometries,
we choose the contour $\gamma=[0,1]$ as the generator of the homology 
group, and the associated canonical form is
\begin{equation}
\Omega(\gamma)=\left(\frac{1}{u}+\frac{1}{1-u}\right)du\,.
\end{equation}
In the terminology of this section, this means
we have a one-dimensional period matrix
\begin{equation}\label{eq:tempEulerBeta1}
	P(\gamma;\Omega(\gamma);\Phi)=\frac{a_1+a_2}{a_1a_2\epsilon}
	\frac{\Gamma(1+a_1\epsilon)\Gamma(1+a_2\epsilon)}
	{\Gamma(1+(a_1+a_2)\epsilon)}=\frac{a_1+a_2}{a_1a_2\epsilon} + \mathcal{O}(\eps^0)\,.
\end{equation}
We will frequently choose bases of the cohomology group that are not the canonical forms of our chosen generators of the homology group. Let us therefore expand this illustrative example by considering a different generator of the cohomology group,
\begin{equation}\label{eq:tempEulerBeta3}
	\varphi=\frac{du}{1-u}\,,
\end{equation}
which happens to be the canonical form of a cycle extending from $u=1$ to infinity.

Keeping the same generator of the homology group, we get
\begin{equation}
	P(\gamma;\varphi;\Phi)=\frac{1}{a_2\epsilon}
	\frac{\Gamma(1+a_1\epsilon)\Gamma(1+a_2\epsilon)}
	{\Gamma(1+(a_1+a_2)\epsilon)} = \frac{1}{a_2\epsilon}+\mathcal{O}(\eps^0)\,.
\end{equation}
Clearly, the two choices are dependent,
\beq
P(\gamma;\Omega(\gamma);\Phi) = 
\frac{a_1+a_2}{a_1}\,P(\gamma;\varphi;\Phi)\,.
\eeq

Let us now compute the intersection of the two choices of forms.
We can use eq.~\eqref{eq:intRes}, which requires the following residues:
\beq
\begin{array}{lll}
 \displaystyle\res_{u=0} \varphi=0\,, \qquad &\qquad 
 \displaystyle\res_{u=1}\varphi=-1\,, \qquad&\qquad 
 \displaystyle\res_{u=\infty}\varphi=1\,, \\
 \displaystyle\res_{u=0} \Omega(\gamma)=1\,, \qquad &\qquad 
 \displaystyle\res_{u=1}\Omega(\gamma)=-1\,, \qquad& \qquad
 \displaystyle\res_{u=\infty}\Omega(\gamma)= 0\,,
 \end{array}
 \eeq
 and, given that $\log \Phi = a_1\eps\log (u)+a_2\eps\log(1-u)$, 
 \beq
 \res_{u=0}\, d\ln\Phi=a_1\eps\,, \qquad \res_{u=1}\,d\ln\Phi=a_2\eps\,,  \qquad
\res_{u=\infty}\,d\ln\Phi= -(a_1+a_2)\eps \,.
\eeq
Then we find that, for instance,
\begin{equation}\label{eq:tempEulerBeta2}
	C(\Omega(\gamma);\Omega(\gamma);\Phi)=
	\frac{a_1+a_2}{a_1a_2\epsilon}\quad\text{and}\quad
	C(\Omega(\gamma);\varphi;\Phi)=\frac{1}{a_2\epsilon}\,.
\end{equation}

We take this opportunity to observe that the poles in $\epsilon$ of
the period matrix arise from logarithmic singularities 
at the boundary of the integration region, which
are regulated by~$\epsilon$. The matrix of intersection
numbers captures the same information (see, e.g.,~ref.~\cite{Mizera:2017rqa}).
For $\gamma=[0,1]$ and when using the form $\varphi$ there is only a contribution 
from the boundary of $\gamma$ at $u=1$, where $\varphi$ also has a pole. When using the canonical form $\Omega(\gamma)$ there is also a contribution from $u=0$.


\section{A coaction on integrals}
\label{sec:coactionOnIntegrals}

Having established our notation in the previous sections, we now 
present the main result of this paper, which is a coaction
$\Delta_\epsilon$ acting on
canonical integrals with a twist as defined in the previous section.

The coaction is given by 
\begin{equation}\label{eq:coactionFormula}
	\Delta_\epsilon\int_{\gamma}\omega=
	\sum_{i,j}\,
	\left[C^{-1}\!\left(\Omega(\vec\gamma);\vec\varphi;\Phi\right)\right]_{ij}
	\int_\gamma\Phi\varphi_i
	\otimes
	\int_{\gamma_j}\omega 
	\,,
\end{equation}
where $\omega=\Phi\varphi$, 
$\vec\varphi\equiv\{\varphi_1,\ldots,\varphi_k\}$ is a set of
differential forms that generate
the cohomology group for the geometry associated to our family of integrals, and 
$\vec\gamma\equiv\{\gamma_1,\ldots,\gamma_k\}$ is a set of
cycles that generate the corresponding homology group. 
The integrands $\varphi$ and cycles $\gamma$ are elements of the cohomology
and homology groups generated by $\vec\varphi$ and $\vec\gamma$.
We stress that the twist $\Phi$ is common to all components in 
eq.~\eqref{eq:coactionFormula}. Furthermore, this coaction
is only valid in the case where the homology group is generated by cells $\gamma_i$, such that for each cell there exists a differential form with logarithmic singularities on the boundary of $\gamma_i$. For example, we may consider that each~$\gamma_i$
is a positive geometry with canonical form $\Omega(\gamma_i)$.

We conjecture that eq.~\eqref{eq:coactionFormula} satisfies
a highly nontrivial relation already stated in eq.~\eqref{eq:coactionLaurentMot},
namely that it is related to the coaction on MPLs,
\beq\label{eq:coactionLaurent}
(L_{\eps}\otimes L_{\eps})\Delta_{\eps} = \Delta L_{\eps}\,,
\eeq
where $L_{\eps}$ is the operator which assigns to a function its Laurent expansion
around $\eps=0$. 
Before verifying this conjecture in a series of examples in the following 
sections, we first make some comments on eq.~\eqref{eq:coactionFormula}.

First, as argued previously, several algebraic properties of a generic
integral are determined by its associated period
matrix, as defined in eq.~\eqref{eq:periodMat}. 
It is thus particularly interesting to study the coaction of the
entries of the period matrix (the coaction of a generic integral then
follows from the relation in eq.~\eqref{eq:genericAsPeriods}). 
Consider eq.~\eqref{eq:coactionFormula} in the special case where 
$\gamma=\gamma_k$ and $\varphi=\varphi_l$.
It then follows that the coaction on the period
matrix  is simply obtained by matrix multiplication:
\begin{equation}\label{eq:coactionPMat}
	\Delta_\eps P_{kl}(\vec\gamma;\vec\varphi;\Phi)=
	\sum_{i,j}
	\left[C^{-1}\!\left(\Omega(\vec\gamma);\vec\varphi;\Phi\right)\right]_{ij}
	P_{ki}(\vec\gamma;\vec\varphi;\Phi)
	\otimes
	P_{jl}(\vec\gamma;\vec\varphi;\Phi)\,.
\end{equation}
Second, one can choose generators of the (co)homology
such that $C\!\left(\Omega(\vec\gamma);\vec\varphi;\Phi\right)=\delta_{ij}$,
in which case the coaction takes a particularly simple form:
\begin{equation}\label{eq:coactionDiagBasis}
	\Delta_\eps P_{kl}(\vec\gamma;\vec\varphi;\Phi)=
	\sum_{i}
	P_{ki}(\vec\gamma;\vec\varphi;\Phi)
	\otimes
	P_{il}(\vec\gamma;\vec\varphi;\Phi)\,.
\end{equation}

Finally, we comment on the relation between the coaction proposed 
here, eq.~\eqref{eq:coactionFormula}, and the one proposed
in ref.~\cite{Abreu:2017enx,Abreu:2017mtm}. The latter relies on 
diagonalizing the generators of the (co)homology group using the semi-simple
projection of the period matrix instead of the matrix of intersection numbers. 
For all examples given in this paper the two procedures give the same
result because of eq.~\eqref{eq:periodIntersections}.
In this paper we prefer the formulation of eq.~\eqref{eq:coactionFormula}
because it is more manifestly symmetric in its treatment of the basis
of contours and differential forms. We also note that our conjecture in eq.~\eqref{eq:coactionFormula} is very reminiscent of the 
formula for the coaction of the tree-level string amplitudes~\cite{Drummond:2013vz,Schlotterer:2012ny,Mafra:2019xms}, with the inverse matrix of intersection numbers being identified with the KLT kernel~\cite{Mizera:2017cqs} and $\eps$ related to the string tension $\alpha'$.

\paragraph{Example: Euler's beta function.} 

We return a last time to the example of Euler's beta function
as an illustration of the application of our coaction formula.
We recall that
\begin{equation}
	\int_{0}^1du\,u^{\alpha_1}(1-u)^{\alpha_2}
	=\frac{\Gamma(1+\alpha_1)\Gamma(1+\alpha_2)}{\Gamma(2+\alpha_1+\alpha_2)}\,,
\end{equation}
where in our case $\alpha_i=n_i+a_i\epsilon$, with $n_i\in\mathbb{Z}$. We take the generator
of the cohomology group to be the canonical form constructed from the integration contour, see eq.~\eqref{eq:basisCohomBeta}.
Using eqs.~\eqref{eq:tempEulerBeta1} and \eqref{eq:tempEulerBeta2} we
find that:
\begin{equation}\label{eq:coactionEulerBeta}
	\Delta_{\epsilon}\left[
	\frac{\Gamma(1+\alpha_1)\Gamma(1+\alpha_2)}{\Gamma(2+\alpha_1+\alpha_2)}
	\right]=
	\frac{\Gamma(1+a_1\epsilon)\Gamma(1+a_2\epsilon)}
	{\Gamma(1+(a_1+a_2)\epsilon)}\otimes
	\frac{\Gamma(1+\alpha_1)\Gamma(1+\alpha_2)}{\Gamma(2+\alpha_1+\alpha_2)}\,.
\end{equation}
Exactly the same result is obtained using the alternative
generator of the cohomology group in eq.~\eqref{eq:tempEulerBeta3}.
It is straightforward to check that eq.~\eqref{eq:coactionLaurent} holds to an arbitrary order in the Laurent expansion 
around $\epsilon=0$.

We can write eq.~\eqref{eq:coactionEulerBeta} explicitly in terms of 
Euler's beta function. If we let $\a=n_a+a\epsilon$, $\b=n_b+b\epsilon$ in eq.~\eqref{eq:betaGenArg}, with $n_a,n_b\in\mathbb{Z}$, then 
\begin{equation}
\label{beta}
\Delta_\eps\left(B(\a,\b)\right) =
\frac{(a\ep)(b\ep)}{(a+b)\ep} B(a\ep,b\ep) \otimes B(\a,\b)\,.
\end{equation}
Finally, we note that eq.~\eqref{eq:coactionEulerBeta} is consistent
with the coaction given in eq.~\eqref{eq:Delta_eps_example_1} for
the gamma function: upon using the fact that the coaction of a product of two 
functions is the product of the coaction of the functions,
$\Delta(f\cdot g)=\Delta(f)\cdot\Delta(g)$, and in particular, that eq.~(\ref{eq:Delta_eps_example_1}) also implies
 \begin{equation}
\Delta\left[\frac{e^{-\gamma_E\eps}}{\Gamma(1+\eps)}\right] 
\,= 
\left[\frac{e^{-\gamma_E\eps}}{\Gamma(1+\eps)}\right] 
\otimes 
\left[\frac{e^{-\gamma_E\eps}}{\Gamma(1+\eps)}\right] \,,
\end{equation}
one may readily verify the coaction in~(\ref{eq:coactionEulerBeta}).


\section{Gauss' hypergeometric function $_{2}F_1$}
\label{sec:2F1}

In this section we discuss  Gauss' hypergeometric function
$_{2}F_1$ in detail. We start from Euler's integral representation given in
eq.~\eqref{2F1} and restrict ourselves to the class of functions
defined below it.
Having already discussed the coaction on beta functions, it
is sufficient to study the parametric integral 
\begin{equation}\label{eq:2f1Simp}
	\int_0^1 u^{n_0+a_0\epsilon} 
	(1-u)^{n_1+a_1\epsilon}
	(1-xu)^{n_{1/x}+a_{1/x}\epsilon} du
	=\frac{\Gamma(\alpha)\Gamma(\gamma-\alpha)}
	{\Gamma(\gamma)}\,_2F_1\left(\alpha,
	\beta;\gamma;x\right)
\end{equation}
where
$\alpha=1+n_0+a_0\eps$,
$\beta=-n_{1/x}-a_{1/x}\eps$, and
$\gamma=2+n_0+n_1+\eps(a_0+a_1)$ and, in accordance with
the framework established in section \ref{sec:construction_of_the_coaction},
$a_0, a_1,a_{1/x}\in\mathbb{C}^*$ and $n_i\in\mathbb{Z}$.
It is well known that for fixed and generic $a_0$, $a_1$ and $a_{1/x}$, 
the space spanned by the parametric integral above is 
two-dimensional (i.e., the cohomology group has dimension 2).
The linear relations between the functions in this space---relations between integrals with different assignments of the integers $n_i$---follow from Gauss' contiguous relations (see, e.g.,~ref.~\cite{handbook}), or equivalently from 
integration-by-parts identities of eq.~\eqref{eq:2f1Simp}.

\subsection{The coaction in the basis of canonical forms}
The integral in eq.~\eqref{eq:2f1Simp} falls into the class of canonical integrals with a twist defined in section~\ref{sec:positiveGeom}. Following the notation
introduced there, we write the integrand of eq.~\eqref{eq:2f1Simp} as
\begin{equation}
	\omega= \Phi \varphi
\end{equation}
where
\begin{align}
\bsp \label{eq:2f1IntTwist}
\Phi &= u^{a_0\eps} (1-u)^{a_1\eps} (1-xu)^{a_{1/x}\eps}\,, \\
\varphi &= u^{n_0} (1-u)^{n_1} (1-xu)^{n_{1/x}}\,du\,.
\esp
\end{align}
The three linear factors 
$P_0(u) = u$, $P_1(u)=1-u$ and $P_{1/x}(u) = 1-xu$ 
define the 0-dimensional hyperplanes
\begin{equation*}
H_0 = \{u=0\}, \quad
H_1 = \{u=1\}, \quad H_{1/x} = \left\{u=1/x \right\},
\end{equation*}
considered to be in general position (i.e., $x\neq0,1,\infty$).
The dimension of the (co)homology group is given by the number
of critical points of $\log\Phi$, i.e., the number of solutions to $d\ln\Phi=0$
in $\mathbb{C}\setminus\{0,1,1/x\}$.
It is straightforward to see
that there are two solutions to this equation, in agreement
with the dimension of the cohomology group deduced from 
Gauss' contiguous relations for the $_2F_1$ function.

It is natural to consider the positive geometries defined by the straight-line segments
\beq\label{eq:2F1_contours}
\gamma_1=[0,1]\qquad \textrm{and} \qquad\gamma_2=[0,1/x]\,,
\eeq 
and the associated canonical forms
\begin{equation}\label{eq:canonicForm2F1}
	\psi_1\equiv\Omega(\gamma_1)=d\log\frac{u}{u-1}\,,
	\qquad
	\psi_2\equiv\Omega(\gamma_2)=d\log\frac{u}{u-1/x}\,,
\end{equation}
as generators of the (co)homology groups of eq.~\eqref{eq:2f1Simp}. 

For a generic integrand $\varphi$ as in eq.~\eqref{eq:2f1IntTwist}, 
we can write the integrals over $\gamma_1$ or $\gamma_2$ in terms of 
Gauss hypergeometric functions:
\begin{align}\bsp
	\int_{\gamma_{1}} \Phi\varphi=&\,
	\frac{\Gamma(\alpha)\Gamma(\gamma-\alpha)}
	{\Gamma(\gamma)}\,_2F_1\left(\alpha,
	\beta;\gamma;x\right)\,,\\
	\int_{\gamma_{2}} \Phi\varphi=&\,x^{-\alpha}
	\frac{\Gamma(\alpha)\Gamma(1-\beta)}
	{\Gamma(1+\alpha-\beta)}
	\,_2F_1\left(\alpha,
	1+\alpha-\gamma;
	1+\alpha-\beta;\frac{1}{x}\right)\,,
\esp\end{align}
with
$\alpha=1+n_0+a_0\eps$,
$\beta=-n_{1/x}-a_{1/x}\eps$, and
$\gamma=2+n_0+n_1+\eps(a_0+a_1)$.
For concreteness we take $0<x<1$, but all results can easily be
analytically continued to any value of $x$. We can then explicitly
construct the period matrix
\begin{equation}\label{eq:periodMat2F1Can}	
	P\!\left(\vec\gamma;
	\vec\psi;\Phi\right)=
	\begin{pmatrix}
	\displaystyle\int_{\gamma_1}\Phi\psi_1 & 
	\displaystyle\int_{\gamma_1}\Phi\psi_2
	\\[3mm]
	\displaystyle\int_{\gamma_2}\Phi\psi_1 & 
	\displaystyle\int_{\gamma_2}\Phi\psi_2
	\end{pmatrix}\,,
\end{equation}
with $\vec\gamma\equiv (\gamma_1,\gamma_2)$ and $\vec\psi = (\psi_1,\psi_2)$. 
We note that all the entries of the period matrix are canonical integrals 
with a twist, as defined in section~\ref{sec:positiveGeom}.

To compute the coaction from eq.~\eqref{eq:coactionPMat} we need the
intersection matrix $C\!\left(\Omega(\vec \gamma);
	\vec\psi;\Phi\right)$
which is easily computed from eq.~\eqref{eq:intRes}:
\begin{equation}\label{eq:matToDiag2F1}	
	C\!\left(\Omega(\vec \gamma);
	\vec\psi;\Phi\right) = C\!\left(\vec \psi;
	\vec\psi;\Phi\right)=
	\begin{pmatrix}
	\dfrac{1}{a_0\epsilon}+\dfrac{1}{a_1\epsilon}  & \dfrac{1}{a_0\epsilon}
	\\[3mm]
	\dfrac{1}{a_0\epsilon}& \dfrac{1}{a_0\epsilon}+\dfrac{1}{a_{1/x}\epsilon} 
	\end{pmatrix}\,.
\end{equation}
By comparing the leading order of the Laurent expansion
of the period matrix with the matrix above we can check that
eq.~\eqref{eq:periodIntersections} holds.

Having at our disposal the matrices $P$ and $C$, we can simply compute the 
coaction by inverting the matrix in eq.~\eqref{eq:matToDiag2F1} and then 
using eq.~\eqref{eq:coactionPMat}. We have done this and checked that it 
satisfies eq.~\eqref{eq:coactionLaurent} by computing explicitly the first 
few terms in the Laurent expansion of all the entries of the period matrix 
in eq.~\eqref{eq:periodMat2F1Can}, and then applying the coaction on the Laurent coefficients expressed in terms of MPLs (the last step was done using the {\sc Mathematica} package {\sc PolyLogTools}~\cite{Duhr:2019tlz}).

\subsection{Coaction in an orthonormal basis}
\label{sec:2f1Orthonormal}

The expression for the coaction obtained in the previous section is not 
particularly elegant, mainly because the matrix in 
eq.~\eqref{eq:matToDiag2F1} (or more precisely, its inverse) is not 
so simple. Here, we show how to obtain a more elegant 
coaction by choosing a different basis of the cohomology group.

We find it convenient to choose basis elements $\varphi_i$ such that 
the matrix of intersection numbers $C(\vec\varphi;\vec\psi;\Phi)$ 
has a minimum number of nonvanishing off-diagonal elements.
This principle is not necessarily compatible with the choice of using the canonical
forms associated with positive geometries. Indeed,
$C$ will be diagonal if each $\varphi_i$ is taken to be a $d\log$-form whose
singularities overlap with the boundary components of  $\gamma_j$ if and only 
if $i=j$. 
However, canonical forms have nonvanishing residues on all boundary components,
which implies that overlaps occur between different basis elements.
In the following we present a basis in which~$C$ is diagonal.
We stress nevertheless that this choice is only a matter of
preference.

For the example at hand, we can choose a form 
$\tilde\varphi_1$ to be singular only at 
$H_1$, and a form $\tilde\varphi_2$ to be singular only at $H_{1/x}$. 
For instance, keeping the original basis of cycles $\gamma_1$ and $\gamma_2$,
we may consider the alternative basis of forms
\begin{equation}\label{eq:froms2f1}
\tilde\varphi_1 = -d\log (1-u) = \frac{du}{1-u}, \qquad 
\tilde\varphi_2 = -d\log(1-xu)=x\frac{du}{1-xu},
\end{equation}
which generate the same cohomology group as the
$\psi_1=\Omega(\gamma_1)$ and $\psi_2=\Omega(\gamma_2)$ 
given in eq.~\eqref{eq:canonicForm2F1}.
With $\tilde{\varphi} = (\tilde\varphi_1,\tilde\varphi_2)$, we find
\begin{equation}
C\!\left(\vec\psi;\tilde\varphi;\Phi\right) = 
\begin{pmatrix}
	\dfrac{1}{a_1\epsilon}  & 0
	\\[3mm]
	0 & \dfrac{1}{a_{1/x}\epsilon} 
	\end{pmatrix}.
\end{equation}
It is then clear that $\vec\varphi=(\varphi_1,\varphi_2)$ with
\begin{equation}\label{eq:cohom2f1}
\varphi_1 = a_1\epsilon \frac{du}{1-u}, \qquad 
\varphi_2 = a_{1/x}\epsilon\,x\,\frac{du}{1-xu},
\end{equation}
produces a matrix of intersection numbers that is unity,
$C\big(\vec\psi;\vec\varphi;\Phi\big)
=\mathbb{1}_2$. When we work in the bases $\vec\gamma$
and $\vec\varphi$ for the (co)homology groups,
eq.~\eqref{eq:coactionDiagBasis} implies the following very compact formula for
the coaction of every element of the associated
period matrix,
\begin{equation}\label{eq:2F1_coac_diag_basis}
	\Delta_\epsilon \int_{\gamma_k}\Phi\varphi_l
	=\int_{\gamma_k}\Phi\varphi_1\otimes\int_{\gamma_1}\Phi\varphi_l
	+\int_{\gamma_k}\Phi\varphi_2\otimes \int_{\gamma_2}\Phi\varphi_l\,.
\end{equation}
We have checked that this coaction agrees with the condition 
\eqref{eq:coactionLaurent}
by computing the full period matrix through order $\epsilon^4$, and verifying
that order by order in $\epsilon$ we reproduce the coaction $\DeltaMPL$ on MPLs.

We finish by writing an explicit formula for the coaction on the Gauss
hypergeometric function $_2F_1$, considering the parametric integral defined in eq.~\eqref{eq:2f1Simp}. Retaining the same bases of cycles $\vec\gamma$ and forms $\varphi$ in eqs.~(\ref{eq:2F1_contours}) and (\ref{eq:cohom2f1}), respectively, where $C\big(\Omega(\vec\gamma);\vec\varphi;\Phi\big)$ is the unit matrix, 
and using the general formula (\ref{eq:coactionFormula}), the coaction takes the form
\bea\label{eq:coaction2f1}
\Delta_{\eps}\left(\int_{\gamma_{1}} \Phi\varphi \right)
= \int_{\gamma_{1}} \Phi\varphi_1 
\otimes \int_{\gamma_{1}} \Phi\varphi
+ \int_{\gamma_{1}} \Phi\varphi_2 
\otimes \int_{\gamma_{2}} \Phi\varphi\,,
\eea
with $\Phi$ and $\varphi$ as in eq.~\eqref{eq:2f1IntTwist}.
To reproduce the right-hand side of 
eq.~\eqref{eq:2f1Simp}, we then substitute
$\alpha=1+n_0+a_0\eps$,
$\beta=-n_{1/x}-a_{1/x}\eps$,
$\gamma=2+n_0+n_1+\eps(a_0+a_1)$.
Finally, using the coaction on Euler's beta 
function in eq.~\eqref{eq:coactionEulerBeta} and the relation 
$\Delta(f\cdot g)=\Delta(f)\cdot\Delta(g)$,
we obtain the coaction on the Gauss hypergeometric function $_2F_1$,
\begin{align}
\nonumber\Delta_\epsilon\Big({}_2F_1(\a,\b;\g;x)\Big) &=
{}_2F_1(1+a\eps,b\ep;1+c\ep;x) \otimes {}_2F_1(\a,\b;\g;x) \\
\label{eq:coaction2F1}&- \frac{b\ep}{1+c\ep}\,
{}_2F_1(1+a\ep,1+b\ep;2+c\ep;x) \\
\nonumber& ~~\otimes \frac{\Gamma(1-\b)\Gamma(\g)}{\Gamma(1-\b+\a)\Gamma(\g-\a)}
x^{1-\a}{}_2F_1\left(\a,1+\a-\g;1-\b+\a;\frac{1}{x}\right)\,,
\end{align}
where $\alpha=n_\alpha+a\eps$, $\b=n_\b+b\eps$ and
$\g=n_\g+c\eps$ (in the notation of eq.~\eqref{eq:2f1Simp},
this means that
$a=a_0$, $b=-a_{1/x}$ and $c=a_0+a_1$).

\subsection{Coaction of a degenerate ${}_2F_1$}

Our motivation for studying hypergeometric functions is that
they appear when evaluating Feynman integrals in dimensional
regularization. In these practical applications, one 
usually encounters non-generic hypergeometric functions,
and it is thus important that one is able to handle these
degenerate cases. 
There are two types of degenerations that can affect the general integrand of 
eq.~\eqref{eq:factorizedIntegrand}. The first type occurs at special values of the external variables $x_j$, at which two or more of the polynomial factors coincide. The second type occurs at special values of the exponents $a_I$ in the twist $\Phi$, where any $a_I=0$ or where $\sum_I a_I=0$, which 
we excluded in section~\ref{sec:construction_of_the_coaction}
because these cases require special treatment in the framework of twisted homology and cohomology.
We expect that both types of degenerations can be handled by taking the corresponding limits of our general coaction formula.
In this section we illustrate this in the context of $_2F_1$, and derive coactions of degenerate cases by taking limits of eq.~\eqref{eq:coaction2F1}.
We close this section by proposing that degeneracies of more general hypergeometric functions can be taken systematically through a detailed analysis of twisted cycles.

\paragraph{Special values of the variable.}
The first degeneration that we consider is when $x$ takes particular values. For $_2F_1(\a,\b;\g;x)$, the particular values are $x=0$, and $x=1$, when the factor $1-xu$ in the Euler integral representation \refE{eq:2f1Simp} combines with one of the others, reducing the number of branch points from four (including infinity) to three.
We first consider $x\to 0$, in which case
\begin{equation}
	{}_2F_1(\a,\b;\g;0)=1\,.
\end{equation}
Taking the same limit on the right-hand side of eq.~\eqref{eq:coaction2F1}
we find that the first term gives $1\otimes 1$ and the second term vanishes, 
reproducing the expected result
\begin{equation}
	\Delta_\epsilon\Big({}_2F_1(\a,\b;\g;0)\Big)
	=1\otimes 1\,.
\end{equation}

A more interesting limit is when we set $x \to 1$, in which case
\begin{equation}
	{}_2F_1(\a,\b;\g;1)=\frac{\Gamma(\gamma)\Gamma(\gamma-\a-\b)}
	{\Gamma(\gamma-\a)\Gamma(\gamma-\b)}\,.
\end{equation}
In this limit, the two terms in  eq.~\eqref{eq:coaction2F1} can be shown to reduce to 
\begin{equation}
	\Delta_\epsilon\left(\frac{\Gamma(\gamma)\Gamma(\gamma-\a-\b)}
	{\Gamma(\gamma-\a)\Gamma(\gamma-\b)}\right)=
	\frac{\Gamma(1+c\eps)\Gamma(1+(c-a-b)\eps)}
	{\Gamma(1+(c-a)\eps)\Gamma(1+(c-b)\eps)}
	\otimes
	\frac{\Gamma(\gamma)\Gamma(\gamma-\a-\b)}
	{\Gamma(\gamma-\a)\Gamma(\gamma-\b)}\,,
\end{equation}
where we recall that $\a=n_\a+a\eps$ and similarly for
$\b$ and $\g$. It is simple to verify that this coaction satisfies 
eq.~\eqref{eq:coactionLaurent} and is consistent with the coaction on 
the gamma function given in eq.~\eqref{eq:Delta_eps_example_1}.

\paragraph{Degenerate exponents.} 
We consider the integrand in the Euler representation \refE{eq:2f1Simp} of $_2F_1(\a,\b;\g;x)$ to be degenerate if any of the three exponents is an integer, or if their sum is an integer.
In terms of  the function $\Phi$ given in \refE{eq:2f1IntTwist}, these are the cases when any of $a_0, a_1, a_{1/x}$ is 0, or when $a_0+a_1+a_{1/x}=0$.
In these cases, $\Phi$ does not vanish nor is it singular at the corresponding point (or the point at infinity, in the case of the sum of the exponents being integer). Note also that it is precisely in these cases that the number of critical points of $\log\Phi$ is less than 2.\footnote{However, it would be wrong to conclude that the dimension of (co)homology is less than 2, as the Morse theory arguments require nondegeneracy.}

Recall that in the notation of \refE{eq:coaction2F1}, 
\begin{equation}
a_0=a, \quad a_1=c-a, \quad a_{1/x}=-b, \quad a_0+a_1+a_{1/x}=c-b.
\end{equation}
So, in terms of $\a, \b, \g,$ the degenerations occur when any of $\a$, $\b$, 
$\g-\a$ or $\g-\b$ is an integer (this can also be seen directly from the integral 
representation in (\ref{2F1}) or its symmetric version upon swapping $\alpha$ and $\beta$).
While in principle we should check the coaction in each of these four cases as 
well as in the cases where more than one of the exponents is an integer,  
here we will simply discuss one such case as an example.

We consider the example of $_2F_1(-\epsilon,1;1-\epsilon;x)$, 
already introduced in  eq.~\eqref{eq:2F1Li}, 
for which  we obtained a coaction by `resumming' the 
Laurent expansion to all orders in $\epsilon$. 
In the notation of
eq.~\eqref{eq:coaction2F1} we have $a=-1, b=0, c=-1$.
It is possible to apply eq.~\eqref{eq:coaction2F1} with $b \neq 0$ and then take the limit $b \to 0$ to recover the correct coaction. 
We then obtain\footnote{In taking the limit $\{\b\to1,b\to0\}$, we note
that $b\epsilon\Gamma(1-\b)= b\epsilon\Gamma(-b\epsilon)\to-1$. Furthermore,
${}_2F_1(m,0,n;x)=1$ for any $m,n$.}
\begin{equation}
	\Delta_\epsilon\Big({}_2F_1(-\epsilon,1;1-\epsilon;x)\Big)=
	1\otimes{}_2F_1(-\epsilon,1;1-\epsilon;x)
	-\frac{x\,\epsilon}{1-\epsilon}{}_2F_1(1-\epsilon,1;2-\epsilon;x)
	\otimes x^{\epsilon}\,.
\end{equation}
Noting that 
\begin{equation}
	\frac{x\,\epsilon}{1-\epsilon}\,{}_2F_1(1-\epsilon,1;2-\epsilon;x)
	=F(\eps,x) = \sum_{n=1}^\infty\epsilon^n\textrm{Li}_n(x),
\end{equation}
with $F(\eps,x)$ given in eq.~\eqref{eq:2F1Li},
we find that the expression we obtained for a generic $_2F_1$, eq.~\eqref{eq:coaction2F1}, 
reduces to eq.~\eqref{eq:Delta_eps_example_2} in this degenerate limit.

While we have only discussed the consistency of our coaction with the degeneration of the exponents on an example, we have checked it on several other cases and believe it to be a general feature. Indeed, the degeneration of exponents can be dealt with through a careful analysis of twisted cycles \cite{AomotoKita}. In this paper, we have considered integrals of the form $\int_\gamma \Phi \varphi$, where $\Phi$ is the multi-valued function introducing the twist. Such integrals can equivalently be written as $\int_\cC \varphi$, where $\cC$ is a twisted cycle, a variant of $\gamma$ that contains a choice of branch for the function $\Phi$ and includes windings around the boundaries of $\gamma$, such that the boundary of $\cC$ is zero. The windings are the key feature that makes it possible to take limits of integer exponents of $\Phi$. In such a limit, the twisted cycle reduces to a contour encircling the corresponding boundary or boundaries. The integral can then be evaluated simply by taking residues. We have confirmed that this procedure agrees with the degenerate coaction formulas we have found for $_2F_1.$


\section{One-dimensional integrals: Lauricella functions $F^{(n)}_D$}
\label{sec:one-dimensional}

In this section we study the class of integrals called the Lauricella 
$F^{(n)}_D$ functions, represented by the one-dimensional integral:
\begin{equation}\label{eq:lauricella}
\int_0^1  u^{\a-1} (1-u)^{\g-\a-1} \prod_{i=1}^n (1-x_i u)^{-\b_i} du
=\frac{\Gamma(\a)\Gamma(\g-\a)}{\Gamma(\g)}
F_D^{(n)}(\a;\b_1,\ldots,\b_n;\g;x_1,\ldots,x_n)\,.
\end{equation}
The cases $n=0$ and $n=1$ correspond to
$F^{(0)}_D(\a;\g)=1$ and $F^{(1)}_D(\a;\b;\g;x)={}_2F_1(\a,\b;\g;x)$. 
Here, we first discuss the case
$n=2$, which corresponds to the Appell $F_1$ function, 
and then consider the case of general $n$.
The construction is a simple generalization of what we have seen in section \ref{sec:2F1} for ${}_2F_1$. 
 The twisted homology and cohomology
groups associated with the Lauricella $F^{(n)}_D$
functions have been studied in ref.~\cite{matsumoto2018relative}.

\subsection{The Appell $F_1$ function}
\label{sec:appellF1single}

The parametric representation of the Appell $F_1$ function is obtained by setting 
$n=2$ in eq.~\eqref{eq:lauricella}:
\begin{equation}
	\int_0^1  u^{\a-1} (1-u)^{\g-\a-1} (1-x u)^{-\b}(1-y u)^{-\b'} du
	=\frac{\Gamma(\a)\Gamma(\g-\a)}{\Gamma(\g)}
	F_1(\a;\b,\b';\g;x,y)\,,
\end{equation}
where we have set $x_1=x$ and $x_2=y$, and
$\b_1=\b$ and $\b_2=\b'$.
To construct its coaction we will follow the same
steps as for the $_2F_1$ function
and start from the parametric integral
\begin{equation}\label{eq:paramF1}
\int_{\gamma_{1}}\Phi\varphi=\int_0^1  u^{n_0+a_0\ep} (1-u)^{n_1+a_1\ep} 
(1-x u)^{n_{1/x}+a_{1/x}\ep}(1-y u)^{n_{1/y}+a_{1/y}\ep} du\,,
\end{equation}
with $\gamma_{1} = [0,1]$ and $a_i\in\mathbb{C}^*$ and $n_i\in\mathbb{Z}$.
In the notation of section \ref{sec:construction_of_the_coaction}, we have
\begin{align}\bsp
\Phi &= u^{a_0\ep} (1-u)^{a_1\ep} 
(1-x u)^{a_{1/x}\ep}(1-y u)^{a_{1/y}\ep} \\
\varphi &= u^{n_0} (1-u)^{n_1} 
(1-x u)^{n_{1/x}}(1-y u)^{n_{1/y}}du\,.
\esp\end{align}
The underlying geometry is determined by the equations
\begin{equation}
P_0(u) = u\,,\quad
P_1(u) = 1-u\,,\quad
P_{1/x}(u) = 1-xu\,,\quad
P_{1/y}(u) = 1-yu\,.
\end{equation}
The hyperplanes are points on the real line:
\begin{equation}
H_0 = \{u=0\}\,,\quad
H_1 = \{u=1\}\,,\quad
H_{1/x} = \{u=1/x\}\,,\quad
H_{1/y} = \{u=1/y\}\,.
\end{equation}
The index of the $P_i$ and the $H_i$ carries the information of
which factor in $\Phi$ it is associated with. 
The dimension of the (co)homology groups can be determined in different ways:
we can either count the solutions to $d\log\Phi=0$, or, alternatively,
count the bounded chambers defined by the hyperplanes above. 
One may also consider the basis of integer-shift relations, which can be derived using integration-by-parts identities.  
In any case we find that the dimension is three. As a basis of of the homology group, 
we can choose the cycles
\begin{equation}\label{eq:cyclesF1}
\gamma_{1} = [0,1]\,,\quad
\gamma_{1/x} = [0,1/x]\,,\quad
\gamma_{1/y} = [0,1/y]\,.
\end{equation}
The index of the cycles carries the information of which hyperplanes
make up its boundary.

The cycles $\gamma_{i}$ are positive geometries with canonical forms $\Omega(\gamma_{i})$ 
(see eq.~\eqref{eq:conventionsCan}).
The forms $\Omega(\gamma_{1})$, $\Omega(\gamma_{1/x})$ and
$\Omega(\gamma_{1/y})$ are a basis of the cohomology group. 
However, building on the experience of the $_2F_1$ function, we prefer
to choose an orthonormal basis that generalizes that of section
\ref{sec:2f1Orthonormal}. For the present example, we thus choose the forms
\begin{equation}\label{varphiF1single}
\varphi_{1} =  a_1\epsilon\,\frac{du}{1-u} \,,\quad
\varphi_{1/x} =  a_{1/x} x \epsilon\,\frac{du}{1-u x} \,,\quad
\varphi_{1/y} =  a_{1/y} y \epsilon\,\frac{du}{1-u y} \,.
\end{equation}
With this basis of $d\log$-forms and the cycles
$\vec\gamma$ of eq.~\eqref{eq:cyclesF1}, we find that the matrix of intersection
numbers $C(\Omega(\vec\gamma);\vec\varphi;\Phi)$ is the
identity matrix $\mathbb{1}_3$.

Using eq.~\eqref{eq:coactionFormula}, we can then write the coaction 
for the parametric integral in eq.~\eqref{eq:paramF1}:
\begin{equation}\label{eq:coactionF1}
	\Delta_\eps\int_{\gamma_1}\Phi\varphi
	=\int_{\gamma_1}\Phi\varphi_{1}\otimes\int_{\gamma_{1}}\Phi\varphi
	+\int_{\gamma_1}\Phi\varphi_{{1/x}}\otimes\int_{\gamma_{1/x}}\Phi\varphi
	+\int_{\gamma_1}\Phi\varphi_{{1/y}}\otimes\int_{\gamma_{1/y}}\Phi\varphi\,.
\end{equation}
Finally, to write the coaction on the Appell $F_1$ functions,
we simply need to rewrite the above integrals in terms of the 
$F_1$ functions and use the coaction on the beta function we have 
established in eq.~\eqref{eq:coactionEulerBeta}. 
Explicitly, we have
\begin{equation}
\Phi\varphi=u^{\a-1} (1-u)^{\g-\a-1}(1-x u)^{-\b}(1-y u)^{-\b'} du\,,\\
\end{equation}
with
\begin{align}
\bsp
	\int_{\gamma_1}\Phi\varphi
	&=\frac{\Gamma(\a)\Gamma(\g-\a)}{\Gamma(\g)}
	F_1(\a;\b,\b';\g;x,y)\,,\\
	\int_{\gamma_{1/x}}\Phi\varphi
	&=x^{-\a}\frac{\Gamma(\a)\Gamma(1-\b)}{\Gamma(1+\a-\b)}
	F_1\left(\a;1+\a-\g,\b';1+\a-\b;\frac{1}{x},\frac{y}{x}\right)\,,\\
	\int_{\gamma_{1/y}}\Phi\varphi
	&=y^{-\a}\frac{\Gamma(\a)\Gamma(1-\b')}{\Gamma(1+\a-\b')}
	F_1\left(\a;\b,1+\a-\g;1+\a-\b';\frac{x}{y},\frac{1}{y}\right)\,,
\esp\end{align}
with the identification
$\a=1+n_0+a_0\eps$,
$\g=2+n_0+n_1+(a_0+a_1)\eps$,
$\b=-n_{1/x}-a_{1/x}\eps$ and
$\b'=-n_{1/y}-a_{1/y}\eps$.
The integrals with the forms $\varphi_i$ in the integrand appearing in the left entries of the coaction (\ref{eq:coactionF1}) are easily obtained as special cases of these expressions.

To check that the coaction in eq.~\eqref{eq:coactionF1} is consistent with
eq.~\eqref{eq:coactionLaurent}, we have computed the full period matrix
$P(\vec\gamma;\vec\varphi;\Phi)$ with $\vec\gamma$ in (\ref{eq:cyclesF1}) and  $\vec\varphi$ in (\ref{varphiF1single}),  and its Laurent expansion through order
$\eps^4$, i.e., through weight 4 (given that
we have chosen a basis of $d\log$-forms, the power of $\eps$
and the transcendental weight are aligned). The computations were done using the {\sc Mathematica} package
{\sc PolyLogTools}~\cite{Duhr:2019tlz}. We then checked that the
coaction on the period matrix in eq.~\eqref{eq:coactionDiagBasis}
satisfies eq.~\eqref{eq:coactionLaurent} at each order in $\eps$.
Furthermore, it is easy to check that the coaction we have constructed
for the Appell $F_1$ functions degenerates to the appropriate coactions
(that of the Gauss hypergeometric function or that of the beta function)
when $x$ or $y$ are set to 0 or 1.

\subsection{Coaction for generic Lauricella $F_D^{(n)}$}

It is straightforward to generalize the analysis of the Appell $F_1$ function 
from its representation as a one-dimensional integral to the generic
Lauricella function $F_D^{(n)}$ as defined in eq.~\eqref{eq:lauricella}.
The multivalued part of the integrand is given by
\begin{equation}
	\Phi = u^{a \ep} \prod_{i=0}^n (1-x_i u)^{c_i \ep}\,,
\end{equation}
where we define $x_0=1$, and $a,c_i\in\mathbb{C}^*$.
The single-valued 1-form is
\begin{equation}
	\varphi = u^{m} \prod_{i=0}^n (1-x_i u)^{n_i} du\,,
\end{equation}
with $m,n_i\in\mathbb{Z}$.
It is easy to see that there are $n+1$ solutions to 
$d\log \Phi=0$ and the dimension of the (co)homology
groups is thus $n+1$.

A basis of the homology group is formed by the cycles $\gamma_i\in\vec\gamma$ with
\begin{equation}
	\gamma_i=[0,1/x_i]\,,\qquad
	i=0,\ldots,n.
\end{equation}
As in the previous example, we choose a basis $\vec \varphi$ of the
cohomology group such that $C(\Omega(\vec\gamma);\vec\varphi;\Phi)$ is the
unit matrix $\mathbb{1}_{n+1}$. Such a basis is given by $\varphi_i\in \vec\varphi$ with
\begin{equation}
	\varphi_i=c_i x_i\eps\frac{du}{1-x_i u}\,.
\end{equation}

It then follows from eq.~\eqref{eq:coactionFormula} that
\begin{equation}\label{eq:coactionLauricella}
	\Delta_\eps\int_{\gamma_0}\Phi\varphi=
	\sum_{i=0}^{n}\int_{\gamma_0}\Phi\varphi_i\otimes\int_{\gamma_i}\Phi\varphi\,.
\end{equation}
To obtain the coaction on the Lauricella functions in (\ref{eq:lauricella}) from this expression,
we use
\begin{equation}
\Phi\varphi=u^{\a-1} (1-u)^{\g-\a-1} \prod_{i=1}^n (1-x_i u)^{-\b_i} du\,,
\end{equation}
with
\begin{align}\bsp\label{eq:masterFnD}
	\int_{\gamma_0}\Phi\varphi=&
	\frac{\Gamma(\a)\Gamma(\g-\a)}{\Gamma(\g)}
	F_D^{(n)}(\a;\b_1,\ldots,\b_n;\g;x_1,\ldots,x_n)\,,\\
	\int_{\gamma_i}\Phi\varphi=&x_i^{-\a}
	\frac{\Gamma(\a)\Gamma(1-\b_i)}{\Gamma(1+\a-\b_i)}\\
	&F_D^{(n)}(\a;\b_1,\ldots,\b_{i-1},1+\a-\g,\b_{i+1},\ldots
	\b_n;1+\a-\b_i;y^{(i)}_1,\ldots,y^{(i)}_n)\,,
	\quad\text{for }i\geq1\,,
\esp\end{align}
where $y^{(i)}_j=x_j/x_i$ if $j \neq i$ and $y^{(i)}_i=1/x_i$,
together with the identifications
$\a=1+m+a\eps$,
$\g=2+m+n_0+(a+c_0)\eps$,
$\b_i=-n_i-c_i\eps$.
It is straightforward to check the consistency of this result with the 
 coactions obtained previously for the beta function,
the Gauss hypergeometric function and the Appell $F_1$ function.


\section{Two-dimensional integrals: Appell functions}
\label{sec:two-dimensional}

In this section we study a family of functions,
the so-called Appell functions $F_1$, $F_2$, $F_3$ and $F_4$.
We have already encountered the $F_1$ function in the previous section,
as it also admits a one-dimensional
integral representation.
We now study it from a different perspective.
All Appell functions depend on two external variables (which we always
denote $x$ and~$y$) and admit a 
two-dimensional integral representation, see e.g.~ref.~\cite{GradRyz}.
The Appell $F_1$ is a special case of the Lauricella $F_D^{(n)}$ function with $n=2$, and it is known that these functions have both one-dimensional and $n$-dimensional integral representations \cite{lauricella}.
This function therefore
provides a natural and simple setting to illustrate the generalization of our discussion 
of previous sections beyond one dimension.

In principle this generalization is straightforward, as our 
framework is in no way restricted
to one-dimensional integrals. 
The underlying geometry is now more involved, because it is no longer defined by points on a line but rather
by lines in $\rproj{2}$. We can, nevertheless, proceed in the same way as for one-dimensional integrals: we construct bases of the associated homology and cohomology  groups,
and the coaction is immediately given by eq.~\eqref{eq:coactionFormula}.
However, there is an interesting aspect related to the choice of bases that we wish to highlight: 
in the examples of the Gauss hypergeometric function and the Lauricella $F^{(n)}_D$ function, it was always straightforward to express all entries in the coaction in terms of the same class of function. 
As we will see in subsequent sections, this remains true for all Appell functions. Indeed---while not always straightforward to achieve in practice---with a properly chosen basis we will be able to express all entries of the period matrix of each Appell function in terms of the same type of function.
That this should be possible in the first place can be understood to be a consequence of the fact that each of the Appell functions can be defined by a particular system of linear differential equations (see section~\ref{sec:F2}).  
A proof of this statement in twisted de Rham theory was given in \cite{mimachinoumi} for the Lauricella functions $F_A, F_B, F_C, F_D$, which include the Appell functions as special cases, and for ${}_{p+1}F_p$.
A suitable choice of bases of forms and cycles will be discussed in each example in turn.

We begin by considering the Appell $F_1$ function written as a two-dimensional integral, which we can explicitly compare with the coaction derived in section~\ref{sec:appellF1single} where we used a one-dimensional representation. We proceed in section~\ref{sec:F3} with the Appell $F_3$ function, which has a more complex geometry. Next, we consider the Appell $F_2$ function in section~\ref{sec:F2}, where expressing the right entries in the coaction in terms of Appell $F_2$ functions requires a somewhat intricate choice of basis; we use the basis of cycles introduced in ref.~\cite{goto2015}. We conclude this section by considering the Appell~$F_4$ function in section~\ref{sec:F4}, where we will use a basis of cycles introduced in ref.~\cite{gotomatsumoto}.

\subsection{The Appell $F_1$ function as a double integral~\label{F1double}}

The Appell $F_1$ function can be written as a two-dimensional integral
as follows:
\begin{align}\bsp\label{eq:f1AsDouble}
\int_0^1dv\int_0^{1-v}du\, u^{\beta-1} v^{\beta'-1}
&(1-u-v)^{\gamma-\beta-\beta'-1}(1-xu-yv)^{-\alpha} = \\
&\qquad\qquad
=\frac{\Gamma(\beta)\Gamma(\beta')\Gamma(\gamma-\beta-\beta')}{\Gamma(\gamma)}
F_1(\alpha,\beta,\beta',\gamma;x,y)\,.
\esp\end{align}
\begin{figure}[b]
\centering
\includegraphics[width=5cm,keepaspectratio=true]{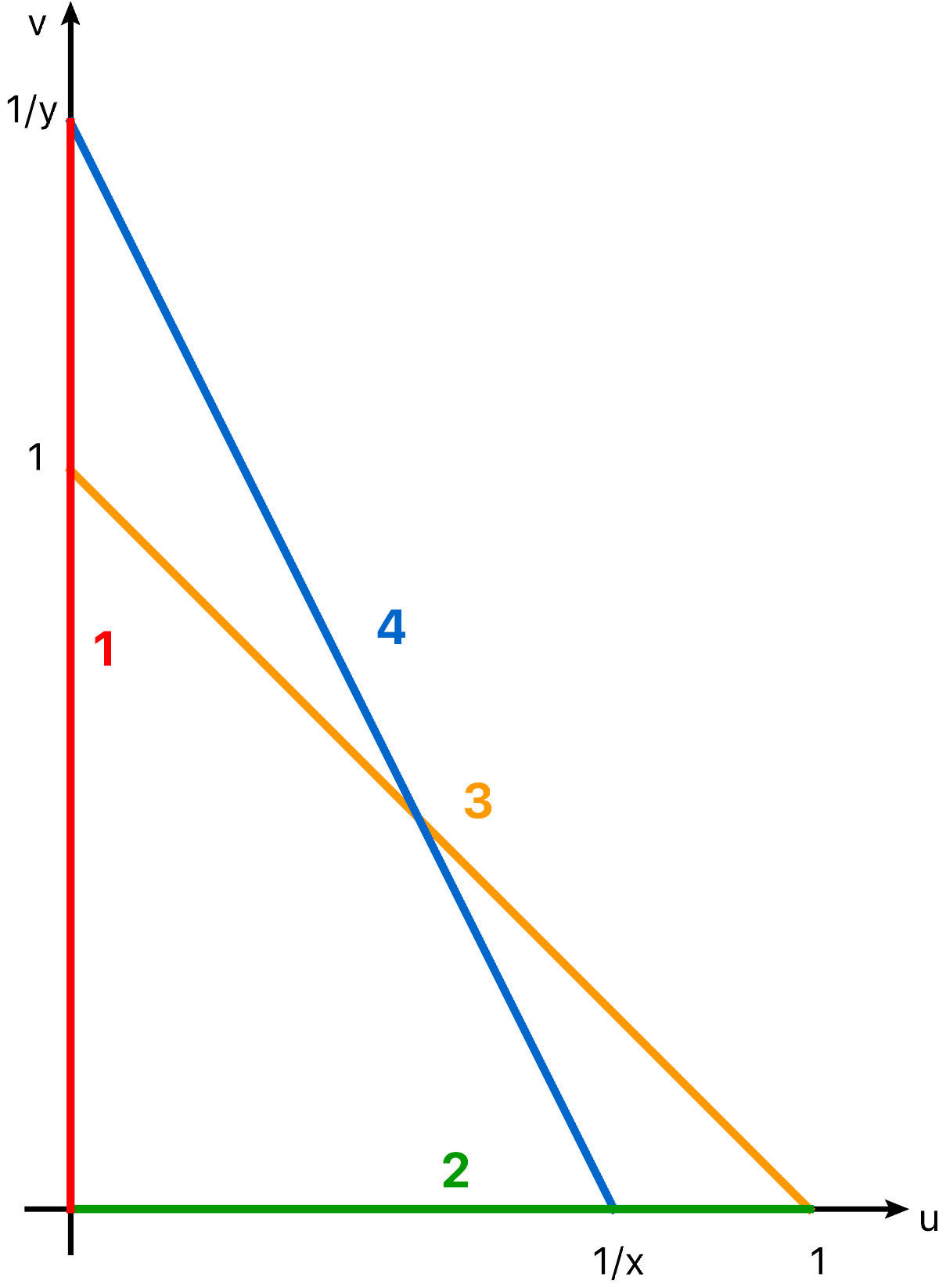}
\caption{Geometry for the two-dimensional integral representation of the
Appell $F_1$ integral of eq.~\eqref{eq:f1AsDouble}}
\label{fig:f1}
\end{figure}
In the notation of section \ref{sec:class_of_integrals}, 
the multivalued part of the integrand is given by
\begin{equation}
	\Phi = u^{c_1\ep} v^{c_2\ep} (1-u-v)^{c_3\ep}(1-xu-yv)^{c_4\ep}\,,
\end{equation}
and the single-valued 2-form is
\begin{equation}
	\varphi=u^{n_1} v^{n_2} (1-u-v)^{n_3}(1-xu-yv)^{n_4} \,du\wedge dv\,.
\end{equation}
The underlying geometry of the parametric integral in eq.~\eqref{eq:f1AsDouble}
is determined by the hyperplanes
\begin{align}
H_1 = \{u=0\},\,\,
H_2 = \{v=0\},\,\,
H_3 = \{1-u-v=0\},\,\,
H_4 = \{1-xu-yv=0\}\,,
\end{align}
which we have represented in fig.~\ref{fig:f1}. It is clear
from this picture that there are three independent
bounded chambers, which means that the dimension of the
second (co)homology groups is 3, as already established in section~\ref{sec:appellF1single}.
We denote by $\gamma_{abc}$ the cycle that is bounded by $H_a$, $H_b$, 
and $H_c$. As a basis we take the following three
triangular regions:
\begin{align}\bsp
\gamma_{123} =& \{0<u<1-v,~~0<v<1\}\,, \\
\gamma_{124} =& \{0<u<(1-yv)/x,~~0<v<1/y\}\,, \\
\gamma_{234} =& \{(1-yv)/x<u<1-v,~~0<v<(x-1)/(x-y)\} \,,
\esp\end{align}
where the inequalities on the right-hand sides are written under the
assumption that, as in fig.~\ref{fig:f1}, $x>1>y>0$.\footnote{
\label{foot:poles}
  We chose this configuration so that the intersection
  of the hyperplanes $H_3$ and $H_4$ is inside the first
  quadrant of the $(u,v)$ plane. This leads to 
  a pole associated with the factor $1-xu-yv$ inside the standard
  integration region, $\gamma_{123}$. Whenever such an issue arises,
  we assume that $x$ and $y$ have small positive imaginary parts so that
  the pole is shifted into the complex plane and the integral is well
  defined.
}
The cycle
$\gamma_{123}$ is the integration contour that appears in the
representation of the $F_1$ function as a double integral, see
eq.~\eqref{eq:f1AsDouble}.

The cycles $\gamma_{abc}$ define positive geometries with associated
canonical forms. For instance,
\begin{equation}
  \Omega(\gamma_{123})=d\ln\frac{u}{1-u-v}\wedge d\ln\frac{1-u-v}{v}
  =-\frac{du\wedge dv}{uv(1-u-v)}\,.
\end{equation}
As in the case of one-dimensional integrals, choosing
$\Omega(\gamma_{123})$, $\Omega(\gamma_{124})$ and 
$\Omega(\gamma_{234})$ as a basis of the cohomology would lead
to a matrix $C(\Omega(\vec\gamma);\Omega(\vec\gamma);\Phi)$ with 
off-diagonal elements. We thus prefer to work with the alternative basis
\begin{align}\bsp\label{eq:formsF1Double}
  \varphi_{13} &=  c_1c_3\eps^2d\log\left(1-u-v\right) \wedge d\log\left(u\right)
  = \frac{c_1c_3\eps^2du \wedge dv}{u(1-u-v)}\,,\\
  \varphi_{14} &= c_1c_4\eps^2 d\log\left(1-xu-yv\right) \wedge d\log\left(u\right) 
  =  \frac{c_1c_4\eps^2y\,du \wedge dv}{u(1-xu-yv)}\,,\\
  \varphi_{34} &= c_3c_4\eps^2 d\log\left(1-u-v\right) \wedge d\log\left(1-xu-yv\right)
  = \frac{c_3c_4\eps^2(y-x) du \wedge dv}{(1-u-v)(1-xu-yv)}\,,
\esp\end{align}
where once again we use the indices of the hyperplanes $H_a$ to label
the differential forms, i.e., the differential form $\varphi_{ab}$ has logarithmic
singularities on the hyperplanes $H_a$ and $H_b$. The normalization
of the $d\log$-forms is such that the intersection matrix
$C(\Omega(\vec\gamma);\vec\varphi;\Phi)$ is the identity matrix $\mathbb{1}_3$.
These normalizations can be easily determined from 
eq.~\eqref{eq:periodIntersections} and the observation that the leading poles
of the period matrix correspond to the points where both integration variables
have an endpoint singularity. To be more concrete, consider the contour
$\gamma_{123}$, which gives rise to potential double endpoint singularities at
$(u,v)=(0,0)$, $(0,1)$ and $(1,0)$ (corresponding to the intersections
of $H_1$ and $H_2$, $H_1$ and $H_3$, and $H_2$ and $H_3$ respectively,
see fig.~\ref{fig:f1}).
It is then easy to see that there is only a double endpoint singularity
for the form $\varphi_{13}$ at $(u,v)=(0,1)$. The singularity is regulated by the
factors $u^{c_1\eps}$ and $(1-u-v)^{c_3\eps}$ in $\Phi$. The normalization
factor $c_1c_3\eps^2$ for $\varphi_{13}$ in eq.~(\ref{eq:formsF1Double})  then guarantees that
\begin{equation}
  \int_{\gamma_{123}}\Phi \varphi_{13} =1+\mathcal{O}(\eps)\,.
\end{equation}
For the forms $\varphi_{14}$ and  $\varphi_{34}$ no double endpoint singularity is generated, and hence, with the normalization factors in eq.~(\ref{eq:formsF1Double}), we have:
\begin{equation}
  \int_{\gamma_{123}}\Phi \varphi_{14} =0+\mathcal{O}(\eps)
  \quad\textrm{and}\quad
  \int_{\gamma_{123}}\Phi \varphi_{34} =0+\mathcal{O}(\eps)\,.
\end{equation}
A similar situation holds for $\gamma_{124}$ and $\gamma_{234}$: for the former a double pole is generated only for the form $\varphi_{14}$ in eq.~(\ref{eq:formsF1Double}), while for the latter only for the form $\varphi_{34}$. 
Alternatively, we can use eq.~\eqref{eq:intnumbersell2} to determine
the matrix of intersection numbers (note, however, that this formula requires the evaluation of the critical points, which are often nontrivial functions
of the $c_i$).

The coaction on the $F_1$ function can then be obtained using
\begin{equation}\label{eq:coactionF1Double}
  \Delta_\eps\int_{\gamma_{123}}\Phi\varphi
  =\int_{\gamma_{123}}\Phi\varphi_{13}\otimes
  \int_{\gamma_{123}}\Phi\varphi
  +\int_{\gamma_{123}}\Phi\varphi_{14}\otimes
  \int_{\gamma_{124}}\Phi\varphi
  +\int_{\gamma_{123}}\Phi\varphi_{34}\otimes
  \int_{\gamma_{234}}\Phi\varphi\,,
\end{equation}
with
\begin{equation}
\Phi\varphi=\,u^{\beta-1} v^{\beta'-1}(1-u-v)^{\gamma-\beta-\beta'-1}
  (1-xu-yv)^{-\alpha}
\end{equation}
and
\begin{align}\bsp\label{eq:basisF1DoubleInt}
  \int_{\gamma_{123}} \Phi\varphi =& 
  \frac{\Gamma(\beta)\Gamma(\beta')\Gamma(\gamma-\beta-\beta')}{\Gamma(\gamma)}
  F_1(\alpha,\beta,\beta',\gamma;x,y)\,,\\
  \int_{\gamma_{124}} \Phi\varphi =\,& x^{-\beta}y^{-\beta'}
  \frac{\Gamma(\beta)\Gamma(\beta')\Gamma(1-\alpha)}{\Gamma(1-\alpha+\beta+\beta')}
  F_1\left(1+\beta+\beta'-\gamma,\beta,\beta',1-\alpha+\beta+\beta';
  \frac{1}{x},\frac{1}{y}\right)\,,\\
  \int_{\gamma_{234}} \Phi\varphi =\,&
  x^{1+\beta'-\gamma}(x-1)^{\gamma-\alpha-\beta}(x-y)^{-\beta'}e^{i\pi\alpha}
  \frac{\Gamma(1-\alpha)\Gamma(\beta')\Gamma(\gamma-\beta-\beta')}
  {\Gamma(1-\alpha-\beta+\gamma)}\\
  &F_1\left(1-\beta,1-\alpha,\beta',1-\alpha-\beta+\gamma;1-x,
  \frac{y(x-1)}{x-y}\right)\,,
\esp\end{align}
with the identifications 
$n_1+c_1\epsilon=\beta-1$,
$n_2+c_2\epsilon=\beta'-1$,
$n_3+c_3\epsilon=\g-\b-\b'-1$ and
$n_4+c_4\epsilon=-\a$.
We note that for this example it was straightforward to write all integrals
in eq.~\eqref{eq:basisF1DoubleInt} as members of the $F_1$ family through
simple changes of variables.
We have explicitly computed the period matrix
$P(\vec\gamma;\vec\varphi;\Phi)$ and checked that the relation in eq.~\eqref{eq:coactionLaurent}
is satisfied order by order in $\epsilon$ through weight 4.

The coaction for the Appell $F_1$ function constructed from the two-dimensional integral
representation in eq.~(\ref{eq:coactionF1Double})  is equivalent to the one we constructed in 
section~\ref{sec:appellF1single} from the one-dimensional representation. 
To be more precise, the period matrix of section \ref{sec:appellF1single},
which we denote $P_{6.1}$, is related to the one computed in this section,
which we denote $P_{7.1}$, as follows:
\begin{equation}\label{eq:pF1rel}
  P_{6.1}=M\cdot P_{7.1}\cdot M^{-1}\,,
\end{equation}
with
\begin{equation}
M = 
\begin{pmatrix}
  1  & 0 & 0 \\
  1 & 0 & -\frac{a}{b} \\
  \frac{b'-a}{b'} & \frac{a}{b'}
  &\frac{a}{b'}
  \end{pmatrix},
\end{equation}
where $a$, $b$ and $b'$ are the coefficients
of $\epsilon$ in $\a$, $\b$ and $\b'$ respectively,
e.g., $\a=n_\a+a\eps$ with $n_\a\in\mathbb{Z}$.
The matrix $M$ can be interpreted as encoding the relation between the bases of integrands used in section \ref{sec:appellF1single} and in the present section. For a given contour, the relations we obtain
from eq.~\eqref{eq:pF1rel} are special cases of the well-known integer-shift relations (see e.g.~ref.~\cite{Bytev:2011ks}).

\subsection{The Appell $F_3$ function\label{sec:F3}}

\begin{figure}[b]
\centering
\includegraphics[width=7cm,keepaspectratio=true]{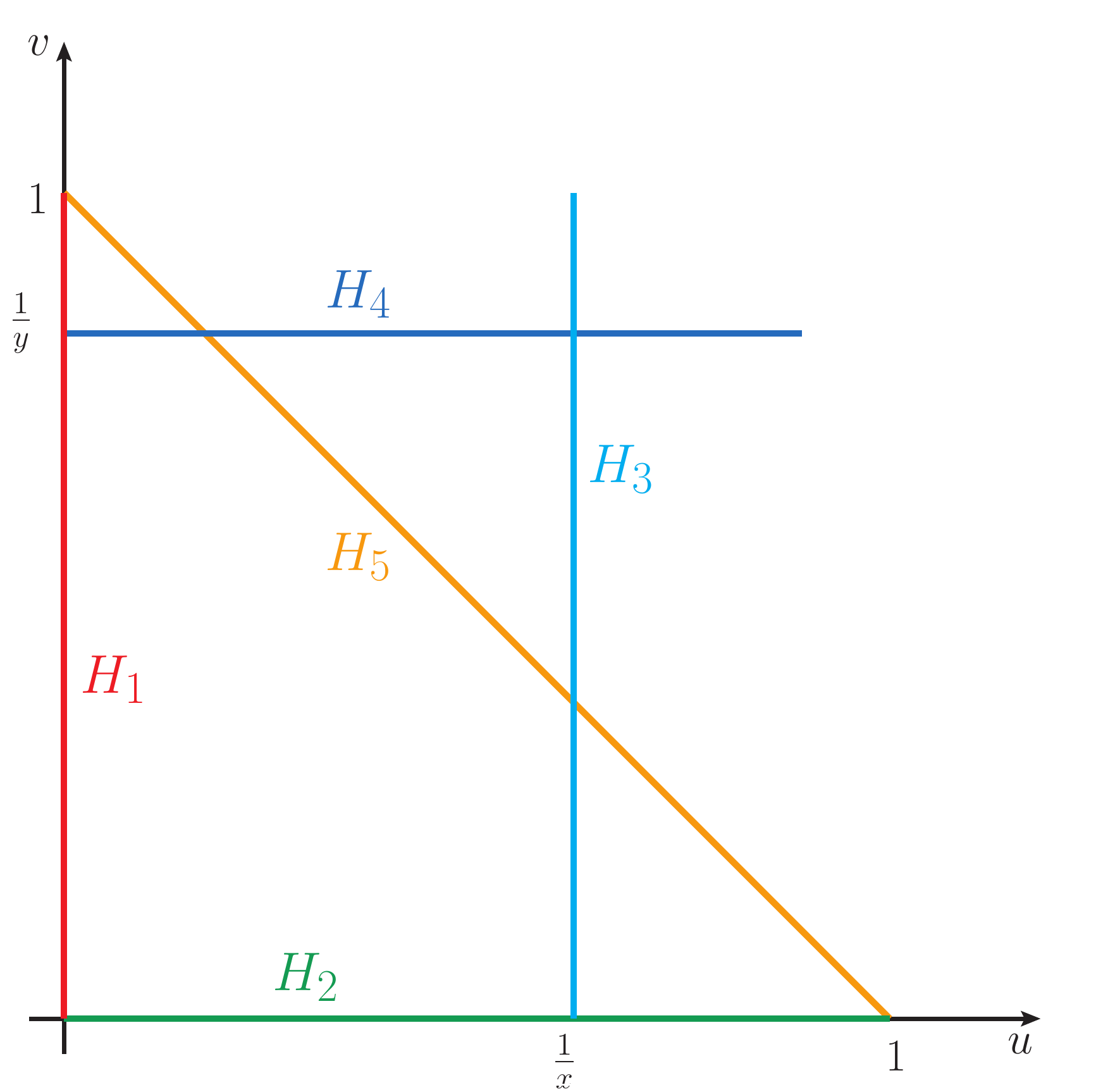}
\caption{Geometry for the Appell $F_3$ integral representation
in eq.~\eqref{eq:F3Def}.}
\label{fig:f3}
\end{figure}

The next example we consider is the Appell $F_3$ function, which
has the following two-dimensional integral representation:
\begin{align}\bsp\label{eq:F3Def}
\int_0^1dv\int_0^{1-v}du\,
u^{\b-1}v^{\b'-1}&(1-u-v)^{\g-\b-\b'-1}(1-xu)^{-\a}(1-yv)^{-\a'}=\\
&\quad=\frac{\Gamma(\b)\Gamma(\b')\Gamma(\g-\b-\b')} {\Gamma(\g)}
F_3(\a,\a',\b,\b',\g;x,y) \,.
\esp\end{align}
This example does not introduce any conceptually new features, but has a slightly more
complicated geometry when compared to the case of the $F_1$ function, as it has
five factors in the integrand.

As usual, we define
\beq\bsp
\Phi &= u^{c_1\ep} v^{c_2\ep} (1-xu)^{c_3\ep} (1-yv)^{c_4\ep} (1-u-v)^{c_5\ep}\,, \\
\varphi &= u^{n_1} v^{n_2} (1-xu)^{n_3} (1-yv)^{n_4} (1-u-v)^{n_5} du \wedge dv\,.
\esp\eeq
The geometry underlying the Appell $F_3$ function is then determined
by the hyperplanes
\begin{align}\bsp
H_1 = &\{u=0\},\quad
H_2 = \{v=0\},\quad
H_3 = \{1-xu=0\},\\
&H_4 = \{1-yv=0\},\quad
H_5 = \{1-u-v=0\},
\esp\end{align}
which we represent in fig.~\ref{fig:f3} for $x>y>1$ {(a similar
comment as the one in footnote \ref{foot:poles} applies)}. The dimension
of the (co)homology groups can be determined by counting the critical
points of $\Phi$ or by counting the independent bounded chambers in
fig.~\ref{fig:f3}. Either way, we find the dimension to be 4.

Similarly to the Appell $F_1$ function case, we
choose triangular cycles $\gamma_{abc}$ defined by the hyperplanes
$H_a$, $H_b$ and $H_c$ as a basis of the homology group. In the present case, we choose $\gamma_{125}$, $\gamma_{235}$, $\gamma_{345}$ and $\gamma_{145}$.
The contour $\gamma_{125}$ is the standard 2-simplex in the definition of the 
$F_3$ function given in eq.~\eqref{eq:F3Def}.
These four cycles are all positive geometries and their canonical
forms provide a basis of the cohomology group.
As in previous examples, we use a different basis, in this case
\begin{align}\bsp
  \varphi_{12} &= c_1c_2\eps^2d\log u \wedge d\log v = 
  \frac{c_1c_2\eps^2du \wedge dv}{uv}\,, \\
  \varphi_{23} &= c_2c_3\eps^2d\log(1-xu)\wedge d\log v = 
  -\frac{c_2c_3\eps^2\,x\,du \wedge dv}{(1-xu)v}\,,\\
  \varphi_{34} &= c_3c_4\eps^2d\log(1-xu) \wedge d\log(1-yv) = 
  \frac{c_3c_4\eps^2\,xy\,du \wedge dv}{(1-xu)(1-yv)}\,,\\
  \varphi_{14} &= c_1c_4\eps^2d\log (u) \wedge d\log(1-yv) = 
  -\frac{c_1c_4\eps^2\,y\,du \wedge dv}{u(1-yv)} \,,
\esp\end{align}
which gives $C(\Omega(\vec\gamma);\vec\varphi;\Phi)=\mathbb{1}_4$.
The normalization of the $d\log$-forms is determined
as described in the $F_1$ example in the discussion below
eq.~\eqref{eq:formsF1Double} above.

It is straightforward to write all the entries of the period matrix 
$P(\vec\gamma;\vec\varphi;\Phi)$ entirely in terms of $F_3$
functions, since we can find changes of variables mapping the arrangement of 
hyperplanes into itself while also mapping $\g_{125}$ into any of the other basis cycles. We thus obtain:
\begin{align}\bsp
&\int_{\g_{125}}\Phi\varphi =
\frac{\Gamma(\b)\Gamma(\b')\Gamma(\g-\b-\b')} {\Gamma(\g)}
F_3(\a,\a',\b,\b',\g;x,y)\\
&\int_{\g_{235}}\Phi\varphi =  (x-1)^{\g-\b-\a}x^{1-\g}e^{i\pi\a}
\frac{\Gamma(1-\a)\Gamma(\b')\Gamma(\g-\b-\b')}{\Gamma(1-\a-\b+\g)}  
\\
&\qquad\qquad\times F_3\left(1-\b,\a',1-\a,\b',1-\a-\b+\g;1-x,\frac{x-1}{x}y\right)
\\
&\int_{\g_{145}}\Phi\varphi =  (y-1)^{\g-\b'-\a'}y^{1-\g}e^{i\pi\a'}
\frac{\Gamma(\b)\Gamma(1-\a')\Gamma(\g-\b-\b')}{\Gamma(1-\a'-\b'+\g)} \\
&\qquad\qquad\times F_3\left(\a,1-\b',\b,1-\a',1-\a'-\b'+\g;\frac{y-1}{y}x,1-y \right)
\\
&\int_{\g_{345}}\!\!\Phi\varphi =
-\frac{x^{\a'+\b'-\g}y^{\a+\b-\g'}e^{i\pi(\g-\b-\b')}}{(x+y-xy)^{\a+\a'+\b+\b'-\g-1}}
\frac{\Gamma(1-\a)\Gamma(1-\a')\Gamma(\g-\b-\b')}{\Gamma(2+\g-\a-\a'-\b-\b')} 
\\
&\times F_3\left(1-\b,1-\b',1-\a,1-\a', 2+\g-\a-\a'-\b-\b';
\frac{x+y-xy}{y},\frac{x+y-xy}{x}\right)
\esp\end{align}
with the identifications
$n_1+c_1\epsilon=\beta-1$,
$n_2+c_2\epsilon=\beta'-1$,
$n_3+c_3\epsilon=-\a$,
$n_4+c_4\epsilon=-\a'$, and
$n_5+c_5\epsilon=\g-\b-\b'-1$.

The coaction on the $F_3$ function then follows from 
eq.~\eqref{eq:coactionFormula}, which yields
\begin{align}\bsp\label{eq:coactionF3}
  \Delta_\eps\int_{\gamma_{125}}\Phi\varphi
  =&\int_{\gamma_{125}}\Phi\varphi_{12}\otimes
  \int_{\gamma_{125}}\Phi\varphi
  +\int_{\gamma_{125}}\Phi\varphi_{23}\otimes
  \int_{\gamma_{235}}\Phi\varphi\\
  &+\int_{\gamma_{125}}\Phi\varphi_{34}\otimes
  \int_{\gamma_{345}}\Phi\varphi
  +\int_{\gamma_{125}}\Phi\varphi_{14}\otimes
  \int_{\gamma_{145}}\Phi\varphi\,.
\esp\end{align}
We have explicitly checked that this coaction satisfies 
the relation in eq.~\eqref{eq:coactionLaurent} for all entries of the period 
matrix $P(\vec\gamma;\vec\varphi;\Phi)$ through weight 4.

The Appell $F_3$ function is the special case with $n=2$ of the Lauricella series $F_B^{(n)}$, and this construction of the coaction generalizes straightforwardly to all $n$.

\subsection{The Appell $F_2$ function\label{sec:F2}}

\begin{figure}[b]
\centering
\includegraphics[width=7cm,keepaspectratio=true]{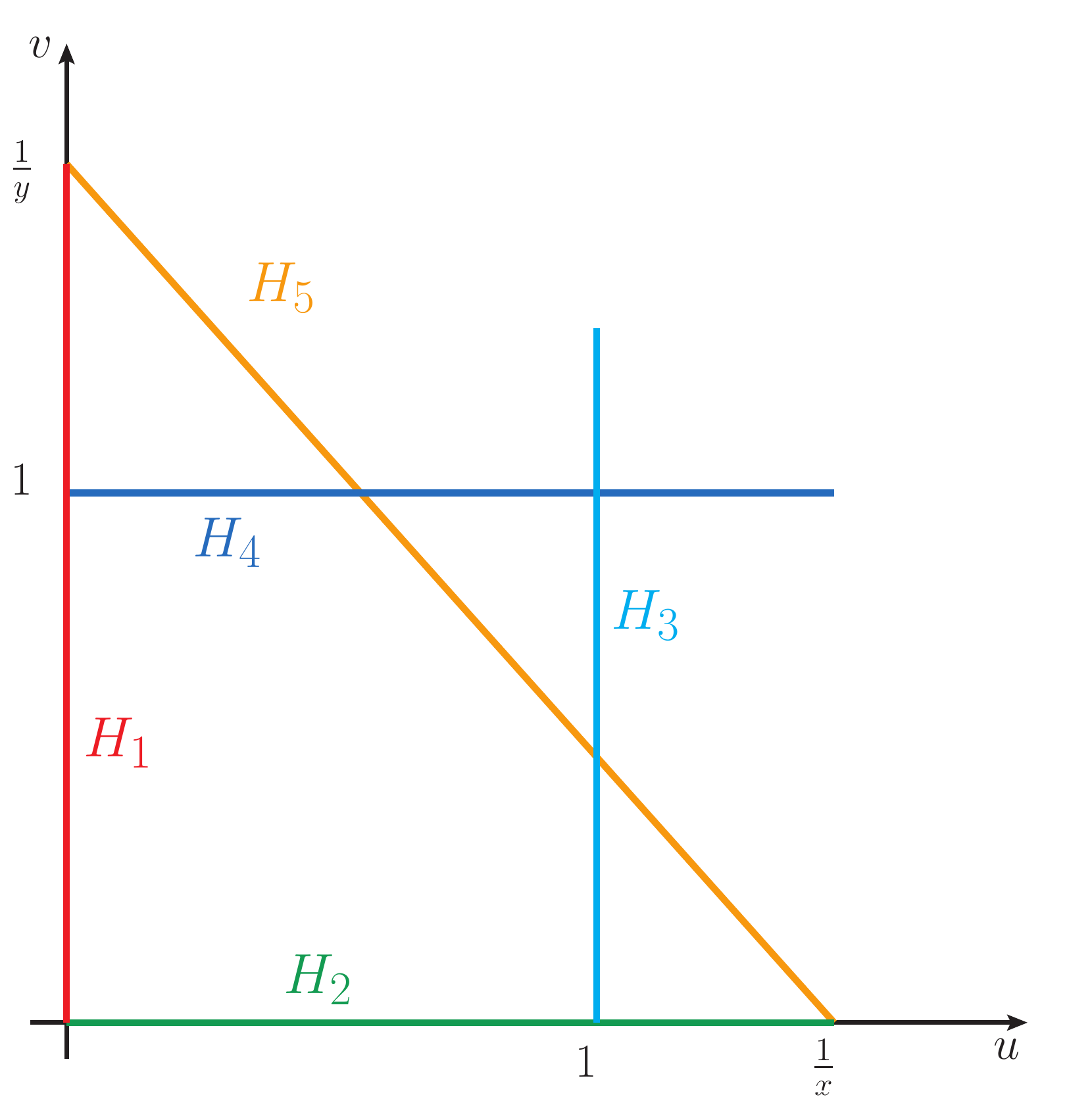}
\caption{Geometry for the Appell $F_2$ integral representation of 
eq.~\eqref{eq:f2Def}}
\label{fig:f2}
\end{figure}

The Appell $F_2$ function can be written as a two-dimensional integral as follows:
\begin{align}\bsp\label{eq:f2Def}
  \int^{1}_0 dv\int^{1}_0 du\,
  u^{\b-1}v^{\b'-1}&(1-u)^{\g-\b-1}(1-v)^{\g'-\b'-1}
  (1-xu-vy)^{-\a}=\\
  &\quad=\frac{\Gamma(\b)\Gamma(\b')\Gamma(\g-\b)\Gamma(\g'-\b')}
  {\Gamma(\g)\Gamma(\g')}
  F_2(\a;\b,\b';\g,\g';x,y) \,.
\esp\end{align}
We follow the same steps as we did for the Appell $F_3$ function, and define
\begin{align}\bsp
\Phi =& u^{c_1\ep} v^{c_2\ep} (1-u)^{c_3\ep} (1-v)^{c_4\ep} (1-ux-vy)^{c_5\ep}\,, \\
\varphi =& u^{n_1} v^{n_2} (1-u)^{n_3} (1-v)^{n_4} (1-ux-vy)^{n_5} du \wedge dv\,.
\esp\end{align}
The underlying geometry is the same as that of the Appell $F_3$ case
studied in the previous section, as can be seen in fig.~\ref{fig:f2}
where we have chosen values of $x$ and $y$ that make the equivalence clear. 
The geometry is determined
by the hyperplanes
\begin{align}\bsp
H_1 = &\{u=0\},\quad
H_2 = \{v=0\},\quad
H_3 = \{1-u=0\},\\
&H_4 = \{1-v=0\},\quad
H_5 = \{1-ux-vy=0\},
\esp\end{align}
which we represent in fig.~\ref{fig:f2} for $0<x,y<1$. It is clear
that there are four bounded chambers, which implies that the dimension
of the (co)homology groups is 4.
We choose the cycles $\gamma_{125}$, $\gamma_{235}$, $\gamma_{345}$ and 
$\gamma_{145}$ as generators of the homology group, where
we recall that the indices correspond to the hypersurfaces $H_i$
that delimit each cycle. This choice does not include the unit square $\gamma_{1234}$
that appears in the definition of the $F_2$ in eq.~\eqref{eq:f2Def}, 
but it can be easily obtained from
\begin{equation}\label{eq:origContF2}
  \gamma_{1234}=\gamma_{125}+\gamma_{345}
  -\gamma_{145}-\gamma_{235}\,.
\end{equation}
As a basis of the cohomology group, we choose the forms
\begin{align}\bsp
  \label{eq:F2forms}
  \varphi_{12} &= c_1c_2\eps^2 d\log (u) \wedge d\log (v) 
  = \frac{c_1c_2\eps^2du \wedge dv}{uv}\,,\\
  \varphi_{23} &= c_2c_3\eps^2 d\log(1-u)\wedge d\log (v) 
  = -\frac{c_2c_3\eps^2du \wedge dv}{(1-u)v}\,,\\
  \varphi_{34} &= c_3c_4\eps^2d\log(1-u) \wedge d\log(1-v) 
  = \frac{c_3c_4\eps^2du \wedge dv}{(1-u)(1-v)}\,, \\
  \varphi_{14} &= c_1c_4\eps^2d\log (u) \wedge d\log(1-v) 
  = -\frac{c_1c_4\eps^2du \wedge dv}{u(1-v)}\,,
\esp\end{align}
which gives $C(\Omega(\vec\gamma);\vec\varphi;\Phi)=\mathbb{1}_4$.

With these definitions, the coaction
on the Appell $F_2$ function can be written as
\begin{align}\bsp\label{eq:coactionF2v1}
  \Delta_\eps\int_{\gamma_{1234}}\Phi\varphi
  =&\int_{\gamma_{1234}}\Phi\varphi_{12}\otimes
  \int_{\gamma_{125}}\Phi\varphi
  +\int_{\gamma_{1234}}\Phi\varphi_{23}\otimes
  \int_{\gamma_{235}}\Phi\varphi\\
  &+\int_{\gamma_{1234}}\Phi\varphi_{34}\otimes
  \int_{\gamma_{345}}\Phi\varphi
  +\int_{\gamma_{1234}}\Phi\varphi_{14}\otimes
  \int_{\gamma_{145}}\Phi\varphi\,.
\esp\end{align}
We explicitly computed the 16 entries of the period matrix
$P(\vec\gamma;\vec\varphi;\Phi)$ and checked that the coaction
\begin{equation}
\label{eq:coaction_PP_F2}
  \Delta_\eps P_{IJ}(\vec\gamma;\vec\varphi;\Phi)=
  \sum_{i}P_{Ii}(\vec\gamma;\vec\varphi;\Phi)
  \otimes
  P_{iJ}(\vec\gamma;\vec\varphi;\Phi)\,.
\end{equation}
satisfies the relation in eq.~\eqref{eq:coactionLaurent} order by order in 
$\epsilon$ through weight 4. 

In all previous examples, we were always able to write all the entries in the coaction
in terms of the same class of function. For example, the coaction of Gauss' hypergeometric function ${}_2F_1$ only involves ${}_2F_1$ functions (and Gamma functions), cf.~e.g.,~eq.~\eqref{eq:coaction2F1}. It is therefore natural to ask if this can still be achieved for the entries in eq.~\eqref{eq:coactionF2v1}, i.e., if all integrals that appear in eq.~\eqref{eq:coactionF2v1} can be written in terms of Appell $F_2$ functions. This is true, 
as shown for example in \cite{mimachinoumi}.
We now present a brief argument for why this must be the case.

Consider the period matrix for the Appell $F_2$ functions. We recall that,
in our conventions (see eq.~(\ref{eq:periodMat})), elements in each row have the same integration contour
and elements in each column have the same integrand 
(without loss of generality, we choose the forms in eq.~\eqref{eq:F2forms}
as generators of the cohomology group). To simplify the argument 
we include $\gamma_{1234}$ in our basis of cycles. 
Then, all elements of the corresponding row are explicitly Appell $F_2$ functions.
Equivalently, in each column there is at least one entry that is explicitly an $F_2$ function.
We now recall that Appell $F_2$ functions satisfy
a given system of second-order differential equations, see e.g.~\cite{GradRyz}. 
For each column (i.e., for each
integrand), there is a different set of differential operators that annihilate the
corresponding $F_2$ function. Since the differential operator is independent of the contour,
it in fact annihilates all elements in the column.
We  conclude that all elements of each column are (linear combinations of) $F_2$ functions, 
since they all satisfy
the Appell $F_2$ differential equations. 
We note that the four elements in each column span the space of solutions of the 
corresponding system of differential equations.

Combined with eq.~\eqref{eq:coaction_PP_F2}, the previous considerations imply that all entries in the coaction can be written in terms of Appell $F_2$ functions. It is, however, not obvious how to achieve this in practice, and we were not able to find any change of variables that allows us to express the integrals in eq.~\eqref{eq:coactionF2v1} in terms of $F_2$ functions. At this point we recall that the integrals depend on the cycles only through their homology classes, i.e., equivalence classes of cycles that differ by boundaries, which integrate to zero. We are thus free to replace the basis of cycles by any other basis for the homology group without changing the space of integrals that they generate.
An alternative basis for the homology group associated to the Appell $F_2$ function was constructed in ref.~\cite{goto2015}. We denote this basis in the following by 
$\vec\Gamma=(
\Gamma_\emptyset,\Gamma_{1},\Gamma_{2},\Gamma_{12}
)$, where $\Gamma_\emptyset=\gamma_{1234}$ as defined in eq.~\eqref{eq:origContF2}.\footnote{We have adapted the notation in ref.~\cite{goto2015}
and set $\Gamma_\emptyset=\Delta$ and $\Gamma_I=\Delta_I$.} 
For details of how the cycles are constructed, we
refer to section~4 of ref.~\cite{goto2015}. Here, we simply
quote the result of Theorem 4.4 of ref.~\cite{goto2015}, which
makes it explicit that the integrals on these cycles
can be written in terms of $F_2$.
Let $\omega$ denote the generic integrand
\begin{equation}
\omega = u^{\b}v^{\b'}(1-u)^{\g-\b-1}(1-v)^{\g'-\b'-1}
(1-x u-y v)^{-\a}\,d\log u \wedge d\log v\,.
\end{equation}
Then, 
\begin{align}\bsp
  \int_{\Gamma_\emptyset}\omega =& 
  \frac{\Gamma(\beta)\Gamma(\gamma-\beta)\Gamma\left(\beta'\right) 
  \Gamma\left(\gamma'-\beta'\right)}
  {\Gamma(\gamma)\Gamma\left(\gamma'\right)}
  F_2\left(\alpha;\beta,\beta';\gamma,\gamma';x,y\right)\,,\\
  \int_{\Gamma_1}\omega =&-e^{i\pi(\beta-\gamma)}x^{1-\gamma}
  \frac{\Gamma(1-\alpha)\Gamma(\gamma -1)\Gamma\left(\beta'\right) 
  \Gamma\left(\gamma'-\beta'\right)}
  {\Gamma(\gamma -\alpha) \Gamma\left(\gamma'\right)}\\
  & \times
  F_2\left(\alpha-\gamma +1;\beta-\gamma+1,\beta';2-\gamma,\gamma';x,y\right)\,,\\
  \int_{\Gamma_2}\omega =&-e^{i\pi\left(\beta'-\gamma'\right)}y^{1-\gamma'}
  \frac{\Gamma(1-\alpha)\Gamma(\beta) \Gamma(\gamma-\beta)
  \Gamma\left(\gamma'-1\right)}
  {\Gamma(\gamma)\Gamma\left(\gamma'-\alpha\right)}\\
  & \times
  F_2\left(\alpha-\gamma'+1;\beta,\beta'-\gamma '+1;\gamma,2-\gamma';x,y\right)\,,\\
  \int_{\Gamma_{12}}\omega =& 
  e^{i\pi\left(\beta'+\beta-\gamma'-\gamma\right)}x^{1-\gamma}y^{1-\gamma'}
  \frac{\Gamma(1-\alpha)\Gamma(\gamma-1)\Gamma\left(\gamma'-1\right)}
  {\Gamma\left(\gamma+\gamma'-\alpha-1\right)}\\
  &\times
  F_2\left(\alpha-\gamma'-\gamma+2;\beta-\gamma+1,\beta'-\gamma'+1;
  2-\gamma,2-\gamma';x,y\right)\,.
\esp\end{align}

Since the cycles in $\vec\Gamma$ are not constructed as linear combinations of bounded chambers,
it is not trivial to relate them to the cycles in $\vec\gamma$. 
However, since they generate the same (twisted) homology group, 
the period matrices $P(\vec\gamma;\vec\varphi;\Phi)$ and
$P(\vec\Gamma;\vec\varphi;\Phi)$ must be related by
a linear transformation of the form
$P(\vec\gamma;\vec\varphi;\Phi)=K\cdot P(\vec\Gamma;\vec\varphi;\Phi)$.
We find that the linear transformation takes the form
\begin{equation}
\label{eq:F2_Delta_gamma}
 K
  =
  \begin{pmatrix}
    \frac{c_3c_4}{(c_1+c_3)(c_2+c_4)}&\frac{c_4}{c_2+c_4}&\frac{c_3}{c_1+c_3}&1\\
    -\frac{c_1c_4}{(c_1+c_3)(c_2+c_4)}&\frac{c_4}{c_2+c_4}&-\frac{c_1}{c_1+c_3}&1\\
    \frac{c_1c_2}{(c_1+c_3)(c_2+c_4)}&-\frac{c_2}{c_2+c_4}&-\frac{c_1}{c_1+c_3}&1\\
    -\frac{c_2c_3}{(c_1+c_3)(c_2+c_4)}&-\frac{c_2}{c_2+c_4}&\frac{c_3}{c_1+c_3}&1
  \end{pmatrix}
  \,.
\end{equation}
Using the change-of-basis matrix $K$,
it is straightforward to rewrite the second entry of each tensor of the coaction
\eqref{eq:coactionF2v1} on the $F_2$ function
in terms of $F_2$ functions, as we have done for all other 
functions discussed above.

The Appell $F_2$ function is the special case with $n=2$ of the Lauricella series $F_A^{(n)}$, and both versions of the construction of the coaction generalize straightforwardly to all $n$. The twisted cycles of ref.~\cite{goto2015} are constructed for arbitrary values of $n$.

\subsection{The Appell $F_4$ function\label{sec:F4}}

For completeness, in this section we discuss the Appell $F_4$ function.
Unlike our previous examples of Appell functions, 
there is no known integral representation with a sufficiently generic description 
in terms of hyperplanes. For example, there is an Euler-type representation with linear factors, given by
\begin{align}\bsp
\label{eq:classicF4}
F_4(\a,\b,\g,\g';&x(1-y),y(1-x)) = 
\frac{\Gamma(\g)\Gamma(\g')}{\Gamma(\a)\Gamma(\b)\Gamma(\g-\a)\Gamma(\g'-\b)}\\
&\times\int_0^1 du \int_0^1 dv\,u^{\a-1}v^{\b-1}(1-u)^{\g-\a-1}(1-v)^{\g'-\b-1}\\
&\times (1-ux)^{\a-\g-\g'+1}(1-vy)^{\b-\g-\g'+1}(1-ux-vy)^{\g+\g'-\a-\b-1}.
\esp\end{align}
However, while the $F_4$ function depends on four parameters $\a$, $\b$, $\g$ and $\g'$,
the integrand has singularities located on 7 hyperplanes, so the
exponents of the different factors are not independent.
Moreover, although the seven factors appearing in this representation represent 
hyperplanes, the arrangement is said to be degenerate, 
because there are points at which three hyperplanes intersect simultaneously
(the points $(u,v)=(0,1/y)$ and $(1/x,0)$, also called non-normal crossings).

To discuss this example along the lines of the previous cases, we could consider a larger class of functions where each of the seven factors in the integrand 
of eq.~\eqref{eq:classicF4} is raised to a different
power, and then take the limit corresponding to the $F_4$ function. This approach
 indeed leads to a coaction on the $F_4$ functions but,
similarly to the case of the $F_2$ family, in this representation it is not obvious that all entries of the coaction are $F_4$ functions.

Instead, we follow a more direct route and use the basis of cycles and
integrands proposed in ref.~\cite{gotomatsumoto}.
The starting point is the Kummer representation
\begin{align}\bsp
\int_{\gamma_1}
&t_1^{\b-\g} t_2^{\b-\g'} L(t_1,t_2)^{\g+\g'-\a-2} Q(t_1,t_2,x,y)^{-\b} dt_1  dt_2=\\
&=\frac{\Gamma(1-\g)\Gamma(1-\g')\Gamma(\g+\g'-\a-1)}{\Gamma(1-\a)}
F_4(\a,\b,\g,\g';x(1-y),y(1-x))\,,
\esp\end{align}
where
\begin{equation}
L(t_1,t_2) = 1-t_1-t_2, \quad
Q(t_1,t_2,x_1,x_2) = t_1 t_2- x(1-y)t_2-y(1-x)t_1 
\end{equation}
and the integration region $\gamma_1$ is a twisted version of the region 
bounded by $L(t_1,t_2)$ and $Q(t_1,t_2,x_1,x_2)$. 
This representation features four hypersurfaces, matching the number of exponents, 
but the polynomial $Q(t_1,t_2,x_1,x_2)$ is quadratic in the $t_i$.
For a precise description of~$\gamma_1$ and the remaining elements of 
the basis of twisted cycles, $\gamma_2$, $\gamma_3$, and $\gamma_4$, 
we refer to ref.~\cite{gotomatsumoto}.\footnote{In that paper, the twisted cycles 
are denoted by $\Delta_i$ rather than $\gamma_i$. We have changed the notation 
in this section to fit with our own conventions.} 
For a generic integrand
\begin{equation}
\omega = t_1^{\b-\g} t_2^{\b-\g'} L(t_1,t_2)^{\g+\g'-\a-2} 
Q(t_1,t_2,x,y)^{-\b} dt_1  dt_2\,
\end{equation}
the integrals over the basis of twisted cycles are then given by \cite{gotomatsumoto}
\begin{align}\bsp\nonumber
&\int_{\gamma_1}\omega=
\frac{\Gamma(1-\g)\Gamma(1-\g')\Gamma(\g+\g'-\a-1)}{\Gamma(1-\a)}
F_4(\a,\b,\g,\g';x(1-y),y(1-x))\,, \\
& \int_{\gamma_2} \omega
=\frac{\Gamma(\a+1-\g)\Gamma(\b+1-\g)\Gamma(1-\b)\Gamma(\g+\g'-\a-1)}
{\Gamma(2-\g)\Gamma(\g')} e^{i\pi(\a+\b-\g-\g')}(x(1-y))^{1-\g}\\
& \qquad \qquad 
F_4(\a-\g+1,\b-\g+1,2-\g,\g';x(1-y),y(1-x))\,,\\
& \int_{\gamma_3} \omega
=\frac{\Gamma(\a+1-\g')\Gamma(\b+1-\g')\Gamma(1-\b)\Gamma(\g+\g'-\a-1)}
{\Gamma(\g)\Gamma(2-\g')} e^{i\pi(\a+\b-\g-\g')}(y(1-x))^{1-\g'}\\
& \qquad \qquad 
F_4(\a-\g'+1,\b-\g'+1,\g,2-\g';x(1-y),y(1-x))\,,\\
& \int_{\gamma_4} \omega
= \frac{\Gamma(\g-1)\Gamma(\g'-1)\Gamma(1-\b)}{\Gamma(\g+\g'-\b-1)} 
(x(1-y))^{1-\g}(y(1-x))^{1-\g'}\\
& \qquad \qquad 
F_4(\a-\g-\g'+2,\b-\g-\g'+2,2-\g,2-\g';x(1-y),y(1-x))\,.
 \esp\end{align}

As a basis of integrands, we adapt the basis of ref.~\cite{gotomatsumoto} 
so that it is explicitly given by $d\log$-forms:
\begin{align}\bsp
\phi_1 &= d\ln\frac{t_1}{L(t_1,t_2)}\wedge d\ln\frac{t_2}{L(t_1,t_2)}
= \frac{dt_1 \wedge dt_2}{t_1 t_2 L(t_1,t_2)}\,,   \\
\phi_2 &= d\ln t_2\wedge d\ln L(t_1,t_2)
= \frac{dt_1 \wedge dt_2}{ t_2 L(t_1,t_2)}\,, \\
\phi_3 &= -d\ln t_1\wedge d\ln L(t_1,t_2)
= \frac{dt_1 \wedge dt_2}{ t_1 L(t_1,t_2)}\,,  \\
\phi_4 &=d\ln\frac{t_1-x}{t_1-1+y}\wedge d\ln\frac{Q(t_1,t_2,x_1,x_2)}{L(t_1,t_2)}
= (1-x-y)\frac{dt_1 \wedge dt_2}{L(t_1,t_2)Q(t_1,t_2,x,y)}\,.
\esp\end{align}
Finally, we define the twist
\begin{equation}
\Phi = 
t_1^{(b-c)\ep} t_2^{(b-c')\ep} L(t_1,t_2)^{(c+c'-a)\ep} 
Q(t_1,t_2,x,y)^{-b\ep}\,,
\end{equation}
and compute the period matrix $P(\vec\gamma;\vec\phi;\Phi)$.
Using eq.~\eqref{eq:periodIntersections} we obtain\footnote{
  {
  We note that the cycles in $\vec\gamma$ are not obviously positive geometries,
  and in principle we do not know how to compute the associated canonical
  forms. Nevertheless, as was done explicitly in the previous section, we 
  can rewrite $\vec\gamma$ in terms of positive geometries
  and then compute $\Omega(\vec\gamma)$. Here we use a shortcut and directly
  compute the matrix of intersection numbers from the period matrix, which is
  possible because we use $d\log$ forms as generators of the cohomology group. 
  }
}
\begin{equation}
  \!\!\!C(\Omega(\vec\gamma);\vec\phi;\Phi)\!
  =\frac{1}{\epsilon^2}\begin{pmatrix}
  \frac{a}{c c' \left(a-c-c'\right)} & \frac{1}{c' \left(a-c-c'\right)} & 
  \frac{1}{c\left(a-c-c'\right)} & \frac{a}{c  c' \left(a-c-c'\right)} \\
  \frac{1}{(c-b) \left(c+c'-a\right)} & 0 & \frac{c'}{(c-a) (c-b) 
  \left(a-c-c'\right)}& \frac{1}{b  \left(a-c-c'\right)} \\
  \frac{1}{\left(c'-b\right) \left(c+c'-a\right)} & \frac{c}{\left(c'-a\right)
  \left(c'-b\right) \left(a-c-c'\right)} & 0 & \frac{1}{b \left(a-c-c'\right)} \\
  \frac{1}{c c'} & 0 & 0 & \frac{b-c-c'}{b c c'}
  \end{pmatrix}.
\end{equation}
We can then construct a new basis of forms $\vec\varphi$ from
$\vec\varphi=\vec\phi^{\,T}\cdot
  C^{-1}(\Omega(\vec\gamma);\vec\phi;\Phi)$,
such that
$P(\vec\gamma;\vec\varphi;\Phi)=\mathbb{1}_4+\mathcal{O}(\epsilon)$.
We do not write an explicit expression for the new basis of forms
$\vec\varphi$ as it is lengthy and not illuminating.

For a generic integrand $\omega=\Phi\varphi$ with $\varphi$
in the cohomology group generated by the forms in $\vec\varphi$, we then find
that
\begin{align}\bsp\label{eq:coactionF4}
  \Delta_\eps\int_{\gamma_{1}}\omega
  =&\int_{\gamma_{1}}\Phi\varphi_{1}\otimes
  \int_{\gamma_{1}}\omega
  +\int_{\gamma_{1}}\Phi\varphi_{2}\otimes
  \int_{\gamma_{2}}\omega
  +\int_{\gamma_{1}}\Phi\varphi_{3}\otimes
  \int_{\gamma_{3}}\omega
  +\int_{\gamma_{1}}\Phi\varphi_{4}\otimes
  \int_{\gamma_{4}}\omega\,.
\esp\end{align}
We have checked
that eq.~\eqref{eq:coactionLaurent} is satisfied for all the entries of the period matrix 
$P(\vec\gamma;\vec\varphi;\Phi)$ through weight 4 by explicit calculation.

The Appell $F_4$ function is the special case with $n=2$ of the Lauricella series $F_C^{(n)}$, but since this is a more complicated positive geometry, it is not so clear how to identify bases suitable for generalizing the construction of the coaction to $n>2$.  The bases of ref.~\cite{gotomatsumoto} are given specifically for Appell $F_4$.


\section{The generalized hypergeometric function $_{p+1}F_p$}
\label{sec:pp1fp}

The last example that we will explore is a class of integrals related to the hypergeometric functions
$_{p+1}F_p$.
This function has a recursive Euler-type integral representation,
\begin{align}\bsp\label{eq:pp1fp}
_{p+1}F_p(\alpha_1,\ldots,&\alpha_{p+1};\beta_1,\ldots,\beta_p;x)=
\frac{\Gamma(\beta_p)}{\Gamma(\alpha_{p+1})\Gamma(\beta_p-\alpha_{p+1})}\\
&\int_0^1u^{\alpha_{p+1}-1}(1-u)^{\beta_p-\alpha_{p+1}-1}{}
_pF_{p-1}(\alpha_1,\ldots,\alpha_p;\beta_1,\ldots,\beta_{p-1};xu)du,
\esp\end{align}
beginning with the trivial case $_1F_0(\alpha;x)=(1-x)^{-\alpha}$.

To construct the coaction on the hypergeometric function $_{p+1}F_p$ we 
consider the differential forms
\begin{equation}
	\omega_p=du_1\wedge\ldots\wedge du_p
	\left(1-x\prod_{i=1}^pu_i\right)^{q+c\epsilon}\prod_{i=1}^p u_i^{n_i+a_i\epsilon}(1-u_i)^{m_i+b_i\epsilon}\,.
\end{equation}
with $n_i,m_i,q\in\mathbb{R}$ and generic $a_i,b_i,c\in\mathbb{C}^*$. We define the twist $\Phi$ in the 
usual way,
\begin{equation}
	\Phi=\left(1-x\prod_{i=1}^pu_i\right)^{c\epsilon}\prod_{i=1}^p u_i^{a_i\epsilon}(1-u_i)^{b_i\epsilon}\,.
\end{equation}
In the following we assume that the range of all products is from 1 to $p$ and do not write it explicitly.
Up to a normalization factor that only involves beta functions, the hypergeometric function $_{p+1}F_p$ 
is given by the integral
\begin{equation}\label{eq:fpdef}
	f_p(x)=\int_{\gamma_{1,\ldots ,p}}\omega_p
\end{equation}
where $\gamma_{1,\ldots, p}=\{(u_1,\ldots,u_p)\in\mathbb{R}^p\,|\,0<u_i<1\}$.

It is easy to determine that the dimensions of the (co)homology groups associated with this
class of integrals is $p+1$. In direct analogy to what was done in the case of the $_2F_1$ 
(see eq.~\eqref{eq:froms2f1}), as a basis of the cohomology group we choose the $d\log$-forms
\begin{align}\bsp
	\phi_{1,\ldots ,p}&=du_1\wedge\ldots\wedge du_p\prod_i\frac{1}{1-u_i}\,,\\
	\phi_{1,\ldots,\hat j,\ldots p,c}&=du_1\wedge\ldots\wedge du_p\frac{1}{1-x\prod_{i}u_i}
	\prod_{i\neq j}\frac{1}{1-u_i}\,.
\esp\end{align}
As a basis of the homology group we choose the cycles
\begin{align}\bsp
	\gamma_{1,\ldots ,p}&=\{(u_1,\ldots,u_p)\in\mathbb{R}^p\,|\,0<u_i<1\}\,,\\
	\gamma_{1,\ldots,\hat j,\ldots ,p,c}&=\left\{(u_1,\ldots,u_p)\in\mathbb{R}^p\,|\,0<u_i<1\,,i\neq j;
	0<u_j<x^{-1}\prod_{i\neq j}u_i^{-1}\right\}\,.
\esp\end{align}
We can then compute the period matrix $P(\vec\gamma;\vec\phi;\Phi)$ to find that
\begin{equation}
	P(\vec\gamma;\vec\phi;\Phi)=
	\frac{1}{\epsilon^p}\mathcal{M}+\mathcal{O}(\epsilon^{p-1})\,,
\end{equation}
where $\mathcal{M}=\text{diag}(d_0^p,\ldots,d_{p}^p)$ with
\begin{equation}
	d_0^p=\prod_i b_i^{-1}\,,\qquad
	d_j^p=\frac{1}{xc}\prod_{i\neq j}
	\frac{a_{j}+b_{j}-a_i-b_i}{b_i(a_{j}+b_{j}-a_i)}
	\quad \text{for }j\geq 1\,.
\end{equation}
Relying on the relation given in eq.~\eqref{eq:periodIntersections}, this determines
the matrix of intersection numbers $C(\Omega(\vec\gamma);\vec\phi;\Phi)$.
It is then trivial to define a new basis $\vec\varphi$ of forms such that 
$C(\Omega(\vec\gamma);\vec\varphi;\Phi)=\mathbb{1}_{p+1}$:
\begin{equation}
	\varphi_{1,\ldots, p}=\frac{\epsilon^p}{d_0^p}\phi_{1,\ldots, p}\,,\qquad
	\varphi_{1,\ldots,\hat j,\ldots ,p,c}= 
	\frac{\epsilon^p}{d_{j}^p}\phi_{1,\ldots,\hat j,\ldots ,p,c}\,.
\end{equation}
As a result, the coaction can be cast in the form
\begin{equation}\label{eq:coactionpp1fp}
	\Delta_\epsilon\int_{\gamma_{1,\ldots, p}}\omega_p
	=\int_{\gamma_{1,\ldots, p}}\Phi\varphi_{1,\ldots, p}\otimes\int_{\gamma_{1,\ldots, p}}\omega_p
	+\sum_{j=1}^p\int_{\gamma_{1,\ldots ,p}}\varphi_{1,\ldots,\hat j,\ldots, p,c}\otimes
	\int_{\gamma_{1,\ldots,\hat j,\ldots ,p,c}}\omega_p\,.
\end{equation}
To obtain the coaction on a $_{p+1}F_p$ function we simply need to normalize the above expression 
by a product of beta functions. One can easily check that this expression reduces to the coaction on the 
$_2F_1$ we constructed in section \ref{sec:2F1} for $p=1$. 
We have also checked explicitly the case $p=2$, corresponding to the $_3F_2$ hypergeometric
function.

We finish by noting that all integrals in the period matrix 
$P(\vec\gamma;\vec\varphi;\Phi)$, and thus in the coaction eq.~\eqref{eq:coactionpp1fp},
can be written in terms of $_{p+1}F_p$ functions \cite{mimachinoumi}. For instance, for $p=2$ this can be done with
the relation
\beq
\bsp
\int_0^1du\int_0^{1/xu}dv&\,\,u^{\alpha_3-1}(1-u)^{\beta_2-\alpha_3-1}v^{\alpha_2-1}
(1-v)^{\beta_1-\alpha_2-1}(1-xuv)^{-\alpha_1}=\\
&\frac{\Gamma(\alpha_2)\Gamma(1-\beta_1)\Gamma(\alpha_3)\Gamma(\beta_2-\alpha_3)}
{\Gamma(\alpha_2-\beta_1+1)\Gamma(\beta_2)}(-1)^{\alpha_2}{}_3F_2(\alpha_1,\alpha_2,\alpha_3;\beta_1,\beta_2;x)\\
&+\frac{\Gamma(1-\alpha_1)\Gamma(\beta_1-1)\Gamma(\alpha_3-\beta_1+1)\Gamma(\beta_2-\alpha_3)}
{\Gamma(\beta_1-\alpha_1)\Gamma(\beta_2-\beta_1+1)}(-1)^{\alpha_2-\beta_1+1}x^{1-\beta_1}\\
&\times{}_3F_2(\alpha_1-\beta_1+1,\alpha_2-\beta_1+1,\alpha_3-\beta_1+1;2-\beta_1,\beta_2-\beta_1+1;x)\,,
\esp
\eeq
which can be generalized to an arbitrary $_{p+1}F_p$.


\section{Summary and discussion}
\label{sec:conclusions}

In this paper we have introduced a coaction $\Delta_{\eps}$ on large classes of hypergeometric functions, for which the coefficients of the Laurent expansion in $\eps$  involve only polylogarithmic functions. In particular, we restrict ourselves to cases where we can find a basis of the homology group associated to the integral such that for each homology generator there is a unique $d\log$-form with singularities on its boundaries. A convenient setting to realize this condition is to consider positive geometries, and all the examples considered in this paper fall into this class. Once an appropriate basis of the (co)homology groups has been identified, we can easily write down the coaction by computing the entries of the period matrix and the matrix of intersection numbers.
The main property of our coaction is that it is consistent with expanding the functions in $\eps$, i.e., acting with $\Delta_{\eps}$ and then expanding each factor in $\eps$ is equivalent to first expanding in $\eps$ and then computing the coaction of the MPLs in the Laurent coefficients. 

We have illustrated our coaction on various hypergeometric functions, in particular on ${}_{p+1}F_p$ and Appell functions, with generalizations to the Lauricella series $F_A$, $F_B$, and $F_D$.  Application to other  hypergeometric functions whose integral representations consist of a product of linear factors raised to generic exponents expanded around integers (hyperplane arrangements) is completely straightforward.  This class includes many of the Lauricella-Saran functions~\cite{saran1955} such as $F_N$ and $F_S$. 

Since many Feynman integrals, at one-loop and beyond, can be expressed in terms of these functions, it will be interesting to connect our results to the recently proposed coaction on Feynman integrals~\cite{Brown:2015fyf,Abreu:2017enx,Abreu:2017mtm}. The coaction on one-loop integrals is by now understood, but the generalization of the results of ref.~\cite{Abreu:2017mtm} beyond one loop is still an open problem. 
The results of this paper can be used to explore the extension of the coaction to two-loop Feynman integrals and beyond, as seen for example in ref.~\cite{Abreu:2018sat}. 

It is important to note that the hypergeometric functions appearing in known Feynman integrals actually violate a key assumption in the results related to twisted (co)homology and intersection numbers, namely that the exponents in the integral representations (the $a_I$ of \eqref{eq:Phiphi}) are nonzero and independent. For this reason, we have emphasized that we are able to derive valid coaction formulas in degenerate limits of the ${}_2F_1$, and
we have argued that a detailed analysis of twisted cycles allows such limits to be taken in general.
Although it was not discussed in this paper, we have checked that we obtain consistent coactions for degenerations of more complicated hypergeometric functions that appear in one-loop integrals. These were all found to be in agreement with the diagrammatic coaction of ref.~\cite{Abreu:2017mtm}. 
We remark that a similar degenerate limit was taken in ref.~\cite{matsubaraheo2019algorithm} in the context of computing intersection numbers in period integrals associated to K3 surfaces.

Our main formula for the coaction is a conjecture, and it would be interesting to prove it rigorously. First steps in this 
direction have recently been taken in ref.~\cite{brown2019lauricella}, albeit in the restricted case of one-dimensional integrals.
Inspired by our preliminary results (see, e.g., conference talks~\cite{Abreu:2018nzy,Abreu:2018sat,ethtalk}), the authors of 
ref.~\cite{brown2019lauricella} have studied in detail the family of Lauricella functions $F_D^{(n)}$ considered in 
section~\ref{sec:one-dimensional}. In particular, they have defined a motivic version of this family of Lauricella functions, and 
they were able to show that it is possible to define a coaction on the Lauricella function $F_D^{(n)}$ which is consistent with 
the (motivic) coaction on MPLs after expansion in $\eps$. The coaction obtained from the motivic setup, however, is not directly 
comparable to the coaction defined here, as we now explain. We have already argued that the second factor in the coaction should 
not change under analytic continuation, and we therefore only consider MPLs modulo their discontinuities in the second entry, 
i.e., only MPLs modulo $2\pi i$. Alternatively, one can interpret the objects in the second entry as single-valued versions of 
hypergeometric functions and MPLs~\cite{Brown:coaction}. This is the approach taken in ref.~\cite{brown2019lauricella}, where all 
Lauricella functions and MPLs in the second entry are the single-valued versions of these functions. It would be interesting to see how this 
alternative choice of 
representing the second entries in the coaction compares to our formula, and in particular what is the role played by the matrix 
of intersection numbers. Here we only mention that the matrix of intersection numbers is closely connected to the computation of 
single-valued functions, as was for example pointed out in the context of string amplitudes~\cite{Mizera:2017cqs,Mizera:2019gea,
Brown:coaction,Brown:2018omk,Brown:2019wna}. It would be fascinating to explore this connection further, and put our conjectured coaction on a rigorous mathematical ground, at least for the case of the Lauricella functions considered in 
ref.~\cite{brown2019lauricella}. 

Our coaction construction is quite general, but we have considered it explicitly on 
certain named hypergeometric functions with well-known integral representations.
Because the second entries of our coaction are constructed by integrating over each of the basis elements of twisted homology, these integrals are not necessarily easy to recognize as belonging to the same class of function. We have argued that they must satisfy the same differential equation, and the methods of refs.~\cite{goto2013twisted,gotomatsumoto,goto2015} may lead to expressions in which this property is manifest, when desired. However, we do not know whether the twisted cycles constructed in these methods have canonical forms of their own. We have discussed this issue in section \ref{sec:F2} for the Appell $F_2$ function, where we were able to deduce a relation between the twisted cycles of ref.~\cite{goto2015} and the homology classes constructed from bounded chambers. It may also be desirable to consider when it is possible to select bases with sparse matrices of intersection numbers, in order to minimize the number of terms in the coaction formula, as we have done for example for ${}_2F_1$. Diagonal matrices of intersection numbers have been constructed for example in ref.~\cite{matsumoto2015pfaffian}.

Other interesting avenues for future research would be to see if the coaction defined here can be extended to more general 
classes of hypergeometric integrals. In particular, here we have restricted ourselves to the study of hypergeometric functions 
whose  expansion in~$\eps$ only involves MPLs. Since, in the motivic setting, the coaction on MPLs is a special case of the 
coaction on  motivic periods (see, e.g., ref.~\cite{Brown:coaction}), it would be interesting to understand if it is 
likewise possible to  extend our  coaction to hypergeometric functions that involve more general periods than MPLs as Laurent 
coefficients.

\acknowledgments
The authors are grateful to Nima Arkani-Hamed, Francis Brown, Clement Dupont, Benjamin Enriquez, Javier Fresan, Riccardo Gonzo, Yoshiako Goto, Martijn Hidding, Saiei-Jaeyong Matsubara-Heo, Sebastian Mizera, Erik Panzer, and Johann Usovitsch for discussions. The authors acknowledge the hospitality of the Galileo Galilei Institute (GGI), Florence, and of the Institute for Theoretical Studies (ITS) of the ETH Zurich during the programmes ``Amplitudes in the LHC era'' and ``Modular Forms, Periods and Scattering Amplitudes.'' SA, CD, EG and JM wish to thank Trinity College Dublin and its Hamilton Mathematics Institute for hospitality, and SA, RB and CD acknowledge the hospitality of the Higgs Center for Theoretical Physics of the University of Edinburgh, at various stages of this work. RB also wishes to thank the Institute for Advanced Study, Princeton, for hospitality. 
EG wishes to thank the CERN theory department for hospitality as a Scientific Associate. This work is supported by the ``Fonds National de la Recherche Scientifique'' (FNRS), Belgium (SA),  by the ERC Consolidator Grant 647356 ``CutLoops'' (RB), the ERC Starting Grant 637019 ``MathAm'' (CD), and the STFC Consolidated Grant ``Particle Physics at the Higgs Centre'' (EG, JM).

\bibliographystyle{JHEP}
\bibliography{bibMain.bib}

\end{document}